%% file: higgsstrahlung-top.tex
\documentclass[11pt]{article}
\usepackage{epsf,amsmath,amssymb,graphicx,scalefnt}
\usepackage{rotating,ifthen,color}
\usepackage{cite}
\usepackage{todonotes,hyperref}
\newcommand{\hw}{W\!H}
\newcommand{\hz}{Z\!H}
\newcommand{\hv}{V\!H}
\newcommand{\gronev}{$V_\text{I}$}
\newcommand{\grtwov}{$V_\text{II}$}
\newcommand{\groner}{$R_\text{I}$}
\newcommand{\grtwor}{$R_\text{II}$}
\newcommand{\dilog}{\text{Li}_2}
\newcommand{\muF}{\mu_\text{F}}
\newcommand{\muR}{\mu_\text{R}}

\newcommand{\higgsstrahlung}{Higgs-Strahlung}
\newcommand{\gfermi}{G_\text{F}}
\newcommand{\mhiggs}{M_{H}}
\newcommand{\mtop}{M_{t}}
\newcommand{\pdf}{{\abbrev PDF}}

\newcommand{\abbrev}{\scalefont{.9}}
\newcommand{\ep}{\epsilon}
\newcommand{\api}{\frac{\alpha_s}{\pi}}
\newcommand{\eqn}[1]{Eq.\,(\ref{#1})}
\newcommand{\fig}[1]{Fig.\,\ref{#1}}

\newcommand{\dd}{{\rm d}}

\newcommand{\order}[1]{{\cal O}(#1)}
\newcommand{\lhc}{{\abbrev LHC}}
\newcommand{\qcd}{{\abbrev QCD}}
\newcommand{\sm}{{\abbrev SM}}

\newcommand{\lo}{{\abbrev LO}}
\newcommand{\nlo}{{\abbrev NLO}}
\newcommand{\nnlo}{{\abbrev NNLO}}

\renewcommand{\Re}{{\rm Re}}

\title{\vspace*{-6em}
  \begin{flushright}
    {\sf\small November 2011 --- CERN-PH-TH/2011-268 ---
    FR-PHENO-2011-016 --- WUB/11-15}
  \end{flushright}
  \vspace*{2em} 
Top-Quark Mediated Effects in Hadronic Higgs-Strahlung}
\author{Oliver Brein$^a$, Robert V. Harlander$^{b}$, Marius
  Wiesemann$^{b}$, Tom Zirke$^b$\\[1em]
{\it $^a$ Physikalisches Institut,
Albert-Ludwigs-Universit\"at Freiburg,}\\
{\it D-79104 Freiburg i.Br,
Germany
}\\[1em]
{\it $^b$Fachbereich C, Bergische Universit\"at Wuppertal,}\\
{\it 42097
  Wuppertal, Germany}
}
\date{}
\begin{document}
\maketitle
\begin{abstract}
Novel contributions to the total inclusive cross section for
\higgsstrahlung{} in the Standard Model at hadron colliders are
evaluated. Although formally of order $\alpha_s^2$, they have not been
taken into account in previous \nnlo{} predictions. The terms under
consideration are induced by Higgs radiation off top-quark loops and
thus proportional to the top-quark Yukawa coupling. At the Tevatron,
their effects to $\hw{}$ production are below 1\% in the relevant Higgs
mass range, while for $\hz{}$ production, we find corrections between
about 1\% and 2\%. At the \lhc{}, the contribution of the newly
evaluated terms to the cross section is typically of the order of
1\%-3\%. Based on these results, we provide updated predictions for the
total inclusive Higgs-Strahlung cross section at the Tevatron and the
\lhc{}.
\end{abstract}



\section{Introduction}

With the \lhc{} experiments becoming sensitive to signals for the
Standard Model (\sm) Higgs boson, the search for this elusive particle
has entered a new and hopefully its final phase.  The direct searches at
{\abbrev LEP}, Tevatron, and {\abbrev LHC} only leave relatively small
allowed windows for the Higgs mass $\mhiggs$, the widest one between
114\,GeV and about 145\,GeV (see Ref.~\cite{atlaslp11,cmslp11} for
preliminary results).

As opposed to evidence or discovery, the exclusion limits rely heavily
on theoretical predictions. The dominant cross section to compare the
experimental measurements to is gluon fusion which receives large
radiative corrections. Although it is probably the most-studied cross
section for an unconfirmed particle, the residual theoretical
uncertainty is still sizable and highly disputed (for a recent
discussion, see Ref.~\cite{Thorne:2011kq}). A lot of this uncertainty is
induced by quantum chromodynamics (\qcd{}), specifically the strong
coupling $\alpha_s$ and the parton density functions (\pdf{}s).

For various reasons, however, gluon fusion need not be the dominant
search mode. The focus of this paper is on the associated production of
a Higgs boson with an electro-weak gauge boson ($pp\to
\hv{},\ V\in\{W^\pm,Z\}$), or ``\higgsstrahlung{}'' for short. At the
Tevatron, where the $\gamma\gamma$ and $b\bar b$-decays of a Higgs boson
produced in gluon fusion cannot be separated from the background with
sufficient precision, this mode is particularly important in the low
mass region. Nominally known through order $\alpha_s^2$, we identify and
evaluate a previously neglected contribution which formally adds to
these next-to-next-to-leading order (\nnlo{}) terms.

For the \lhc{}, the relevance of the \higgsstrahlung{} process used to
be considered marginal. This has changed with the idea of focusing on
events with highly boosted Higgs bosons by analyzing the substructure of
jets~\cite{Butterworth:2008iy}. Even though for a proper theoretical
prediction in such an analysis one needs to consider differential
quantities, it is important to ensure that all effects that contribute
to the total rate are under control.

Once the Higgs mass is known, precise predictions for the individual
production and decay channels will be essential in order to extract the
maximum information from the experiments (see, e.g.,
Ref.~\cite{Zeppenfeld:2000td}).

\begin{figure}
\begin{center}
\begin{tabular}{ccc}
\includegraphics[width=0.31\textwidth]{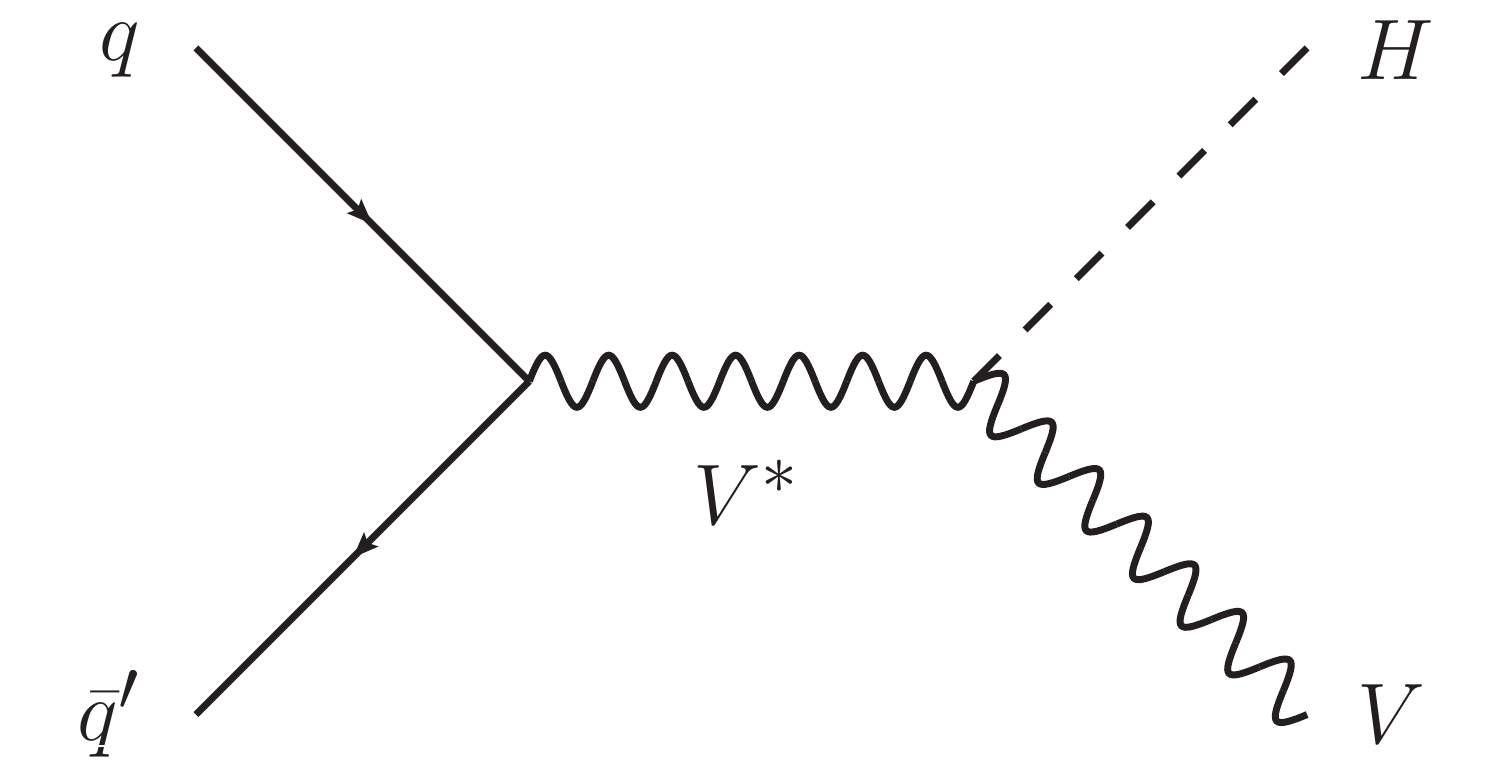} &
\includegraphics[width=0.31\textwidth]{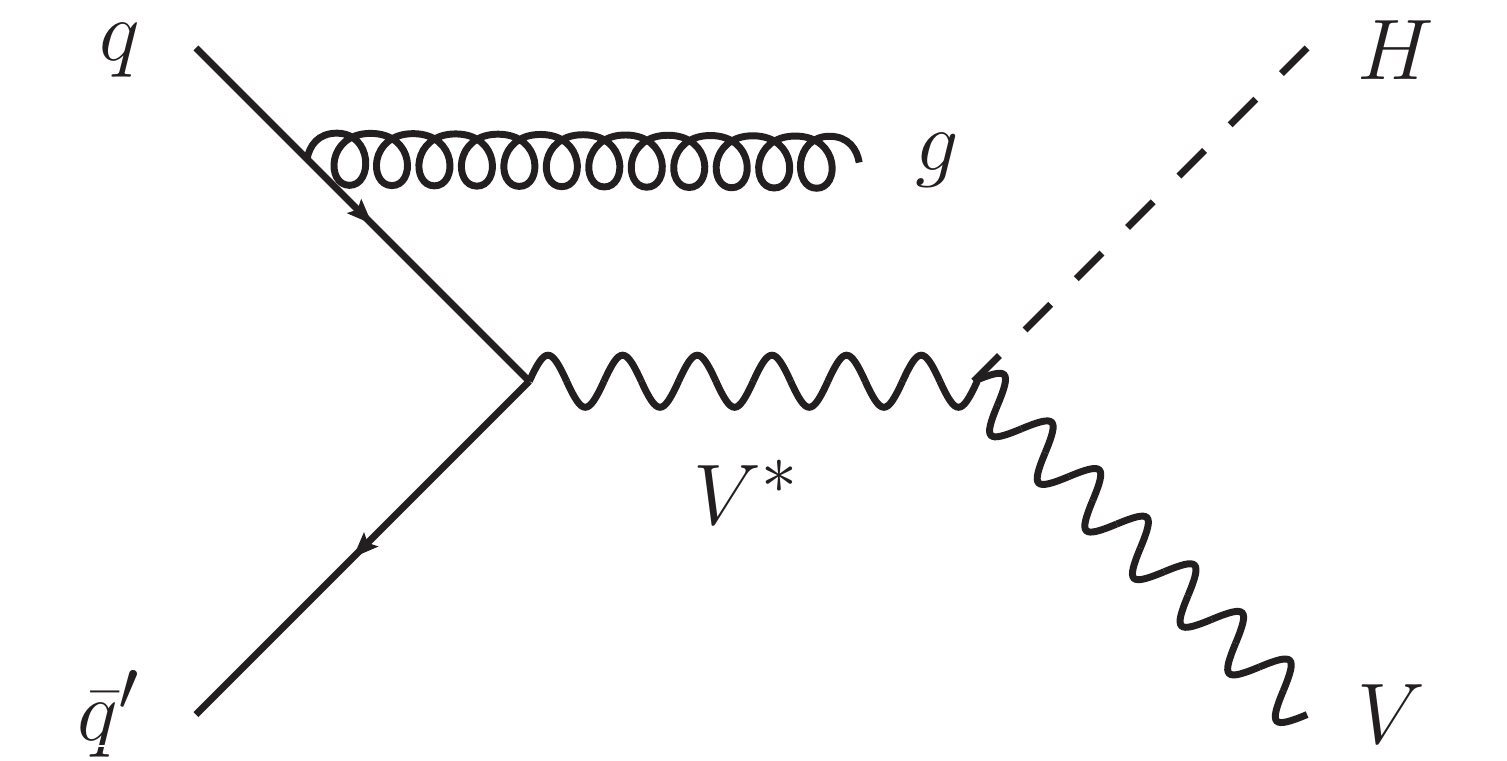} &
\includegraphics[width=0.18\textwidth]{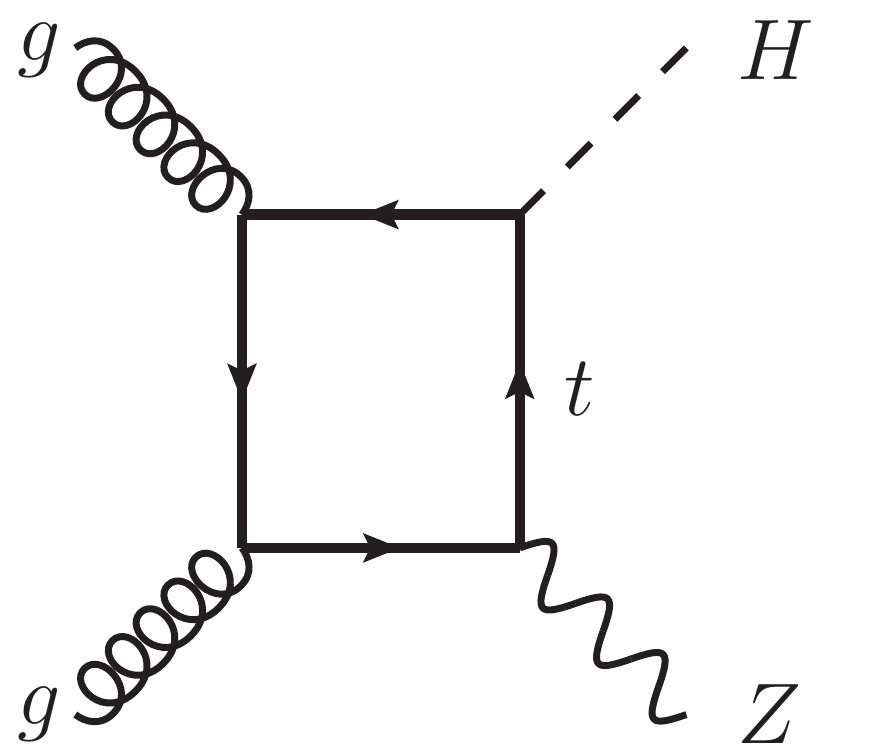} \\
(a) & (b) & (c)
\end{tabular}
\caption{(a) Leading order Feynman diagram contributing to the
  \higgsstrahlung{} process; (b) real corrections at \nlo{} \qcd{}; (c)
  $gg$ component not covered by Drell-Yan-like corrections.}
\label{fig::logg}
\end{center}
\end{figure}

The theoretical prediction of the total inclusive cross section due to
\higgsstrahlung{} at hadron colliders is under very good
control\footnote{For recent work on higher order differential $\hw$
  cross sections, see Ref.~\cite{Ferrera:2011bk}.}: the leading order
contribution is of order $g^4$, where $g$ is the weak coupling constant,
and is completely analogous to what used to be the main search channel
at {\abbrev LEP}, except that the initial $e^+e^-$ is replaced by a
$q\bar q'$ pair, of course, see \fig{fig::logg}\,(a). The \nlo{}
\qcd{}~\cite{Han:1991ia} and the bulk of the \nnlo{} \qcd{}
corrections~\cite{Brein:2003wg}, i.e.\ $\order{g^4\alpha_s}$ and
$\order{g^4\alpha_s^2}$, can be reduced to the Drell-Yan production of a
virtual gauge boson~\cite{Hamberg:1991np,Harlander:2002wh}. The
theoretical uncertainty due to \pdf{}s has been estimated to be at the
percent level, and the renormalization/factorization scale dependence of
these terms is practically negligible~\cite{Dittmaier:2011ti}. For
$\hz{}$ production at $\order{\alpha_s^2}$, however, there are a few
classes of diagrams that have no correspondence to the Drell-Yan
process. For example, the gluon-induced virtual corrections mediated by
a top-quark loop are of order $g^2\lambda^2_t\alpha_s^2$
(see \fig{fig::logg}\,(c)), where $\lambda_t$ is the top-quark Yukawa
coupling which, in the \sm{}, is of order one.  These corrections were
evaluated in Refs.~\cite{Kniehl:1990iv,Brein:2003wg} and found to be of
the order of 5\% at the \lhc{}.

In this paper we consider another class of diagrams which are formally
of order $g^3\lambda_t\alpha_s^2$ and were neglected in previous
analyses. For simplicity, we will refer to them as ``top-mediated terms''
in this paper, even though they are not the only contributions involving
top-quarks, as noted above. Their numerical impact is at the percent
level and therefore within the {\it current} estimated theoretical
uncertainty of the \nnlo{} result (see
Ref.~\cite{Dittmaier:2011ti}). Note, however, that this uncertainty
estimate is dominated by the effects from \pdf{}s and $\alpha_s$; once
these will be known with higher precision, the results of this paper
will be required for the perturbative part to compete with this
precision.

The remainder of this paper is organized as follows:
Section~\ref{sec::calc} defines the effects to be calculated, briefly
describes the methods applied, and presents analytical expressions for
part of the results. In Section~\ref{sec::numerics}, we study the size
of the newly evaluated effects and present updated values for the total
inclusive cross section for $\hw{}$ and $\hz{}$ production at the
Tevatron and the \lhc{} at collision energies of 7 and 14\,TeV.


\section{Calculational details}\label{sec::calc}


\subsection{Outline of the problem}\label{sec::outline}

The Feynman diagrams of the top-mediated terms considered in this paper
can be divided into four groups which will be described in this section.

\begin{figure}
\begin{center}
\begin{tabular}{ccc}
\includegraphics[width=0.26\textwidth]{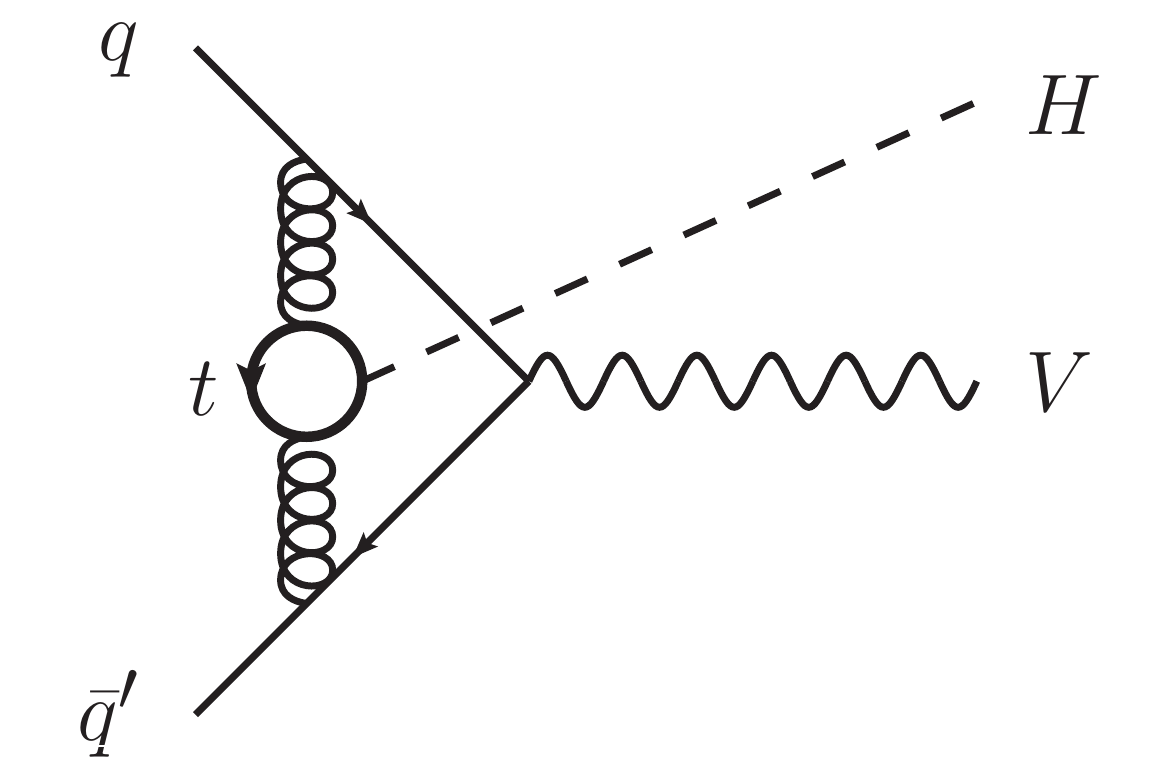} &
\hspace{0.5cm} \includegraphics[width=0.26\textwidth]{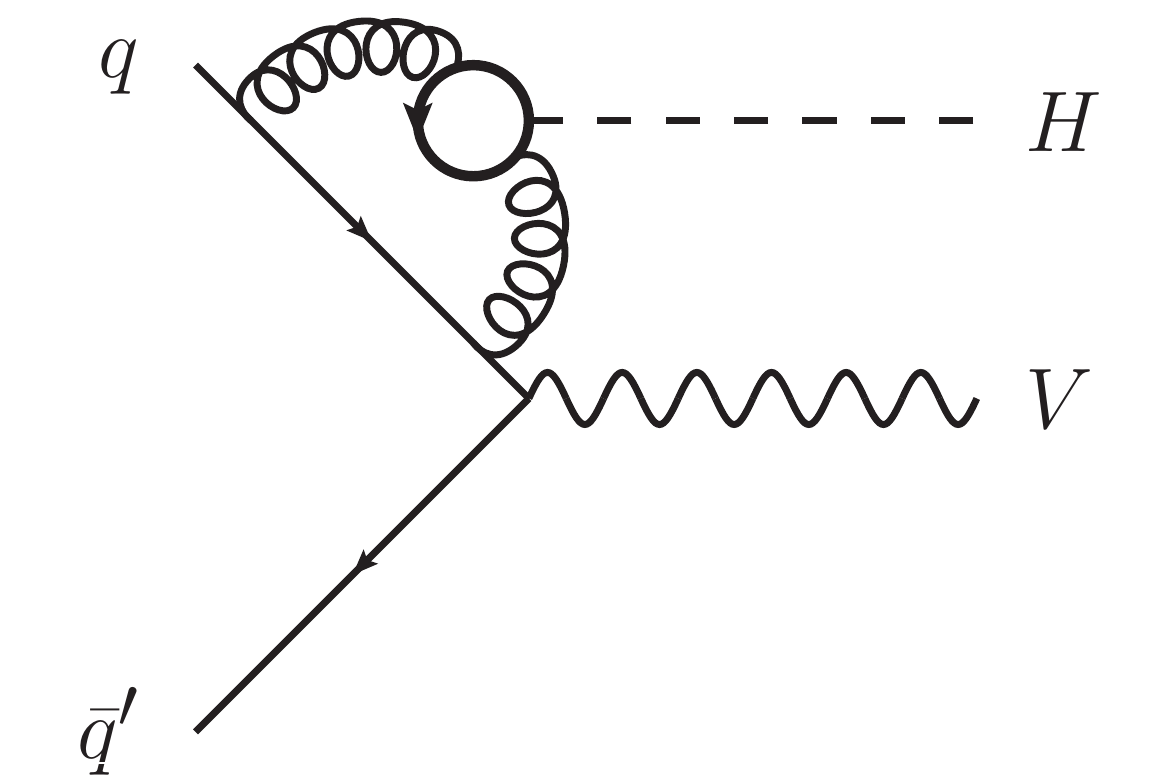} &
\hspace{0.5cm} \includegraphics[width=0.26\textwidth]{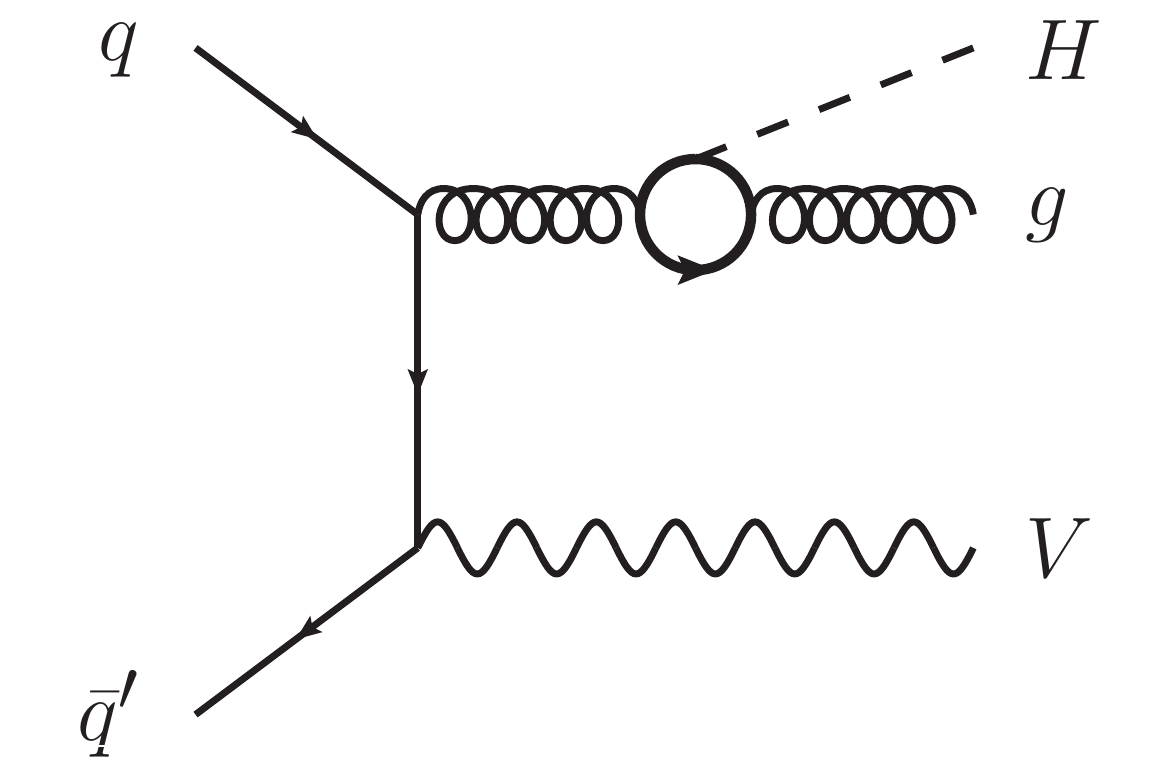} \\
(a) & (b) & (c)
\end{tabular}
\caption{(a),(b) Diagrams of group \gronev{} and (c) group \groner{}
  contributing to the process $q \bar q \rightarrow \hv{}(g)$ at order
  $g^3\lambda_t\alpha_s^2$.}
\label{fig::vhneu}
\end{center}
\end{figure}

\begin{figure}
\begin{center}
\begin{tabular}{ccc}
\includegraphics[width=0.26\textwidth]{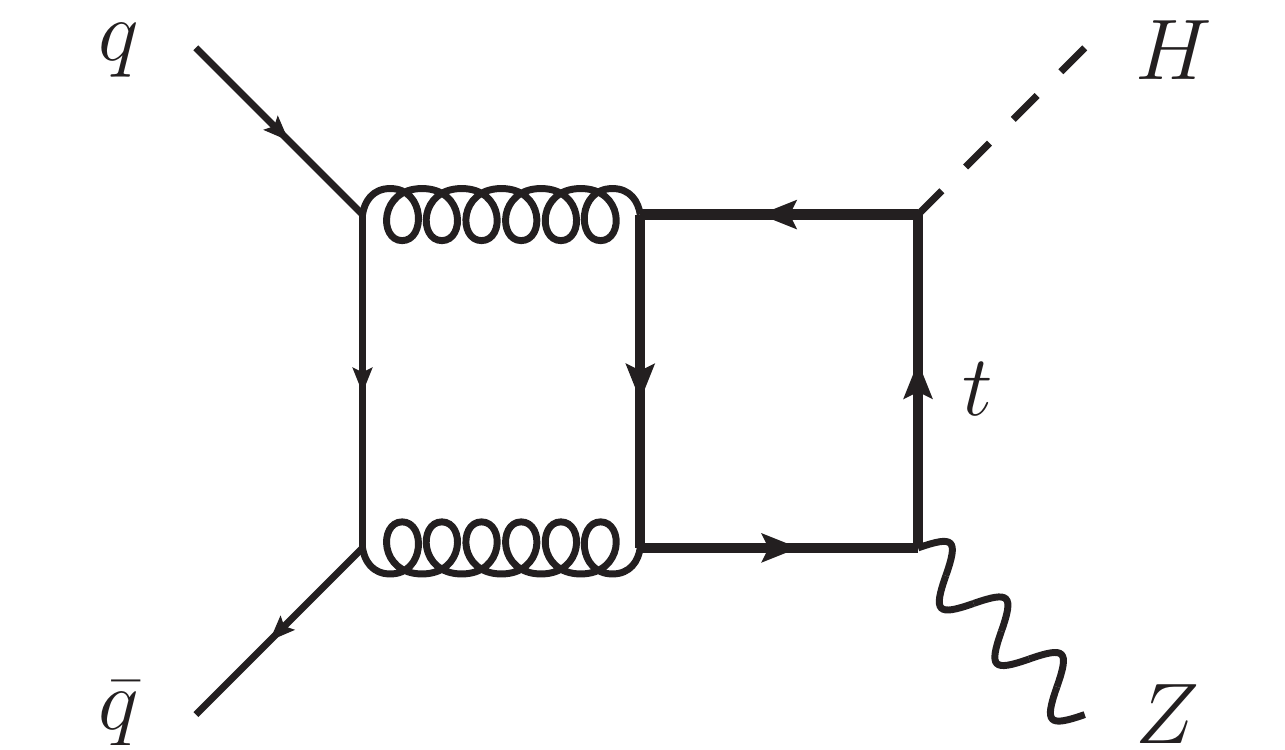} &
\includegraphics[width=0.26\textwidth]{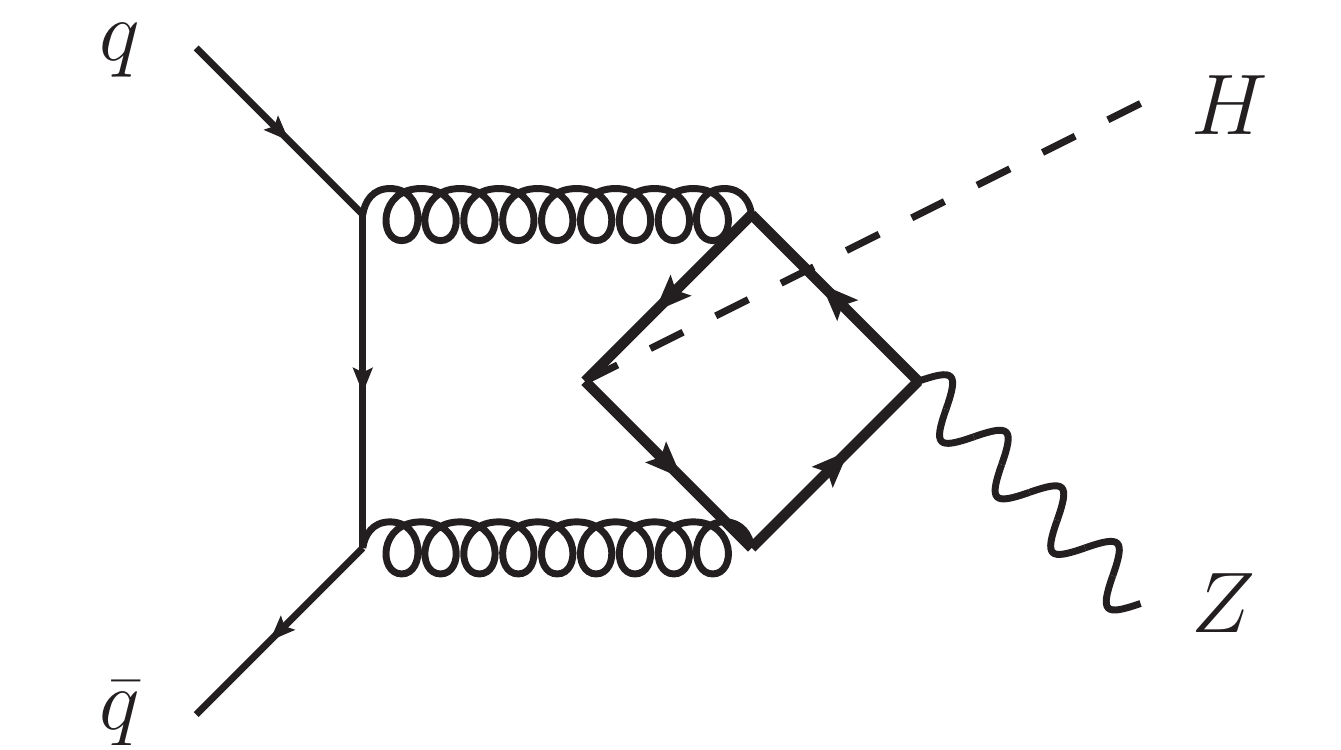} &
\hspace{0.5cm} \includegraphics[width=0.29\textwidth]{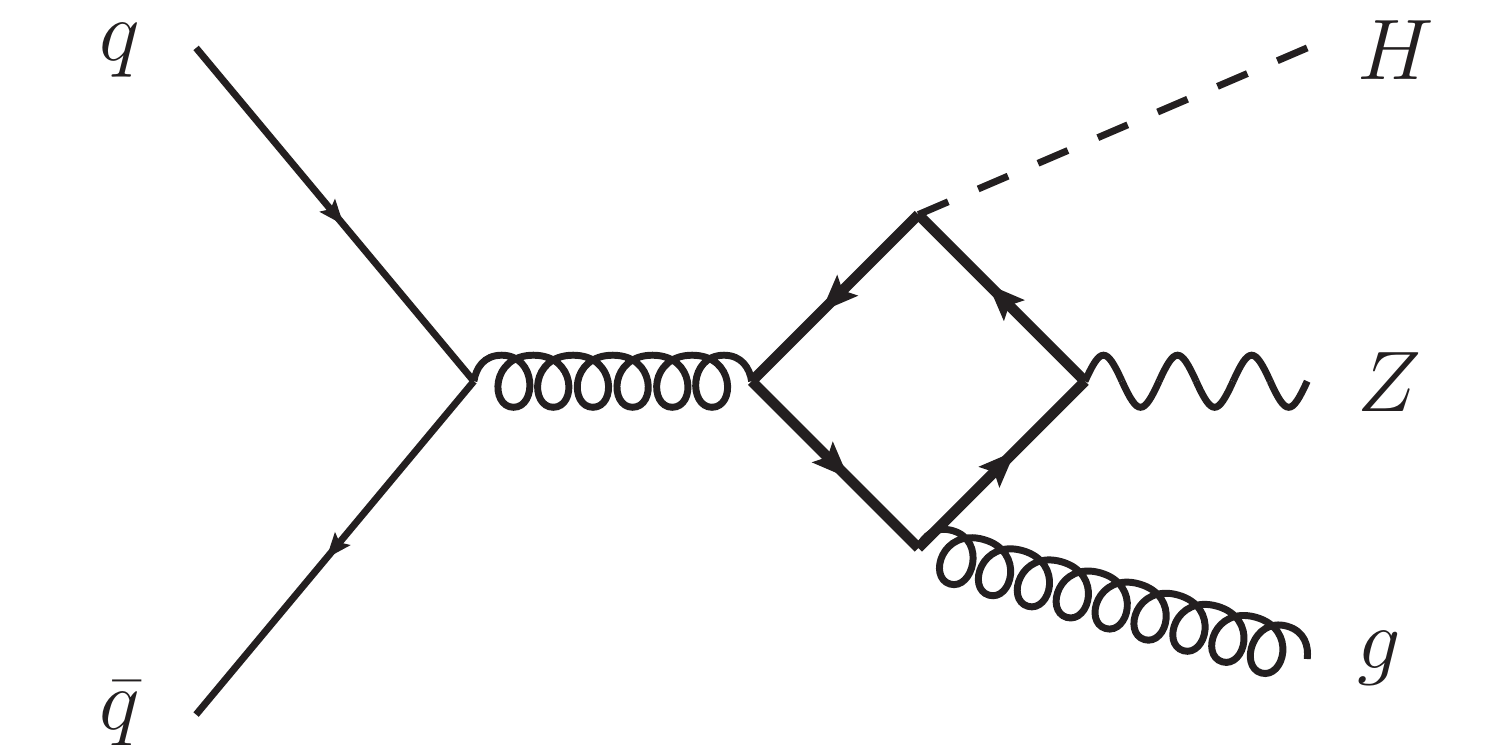} \\
(a) & (b) & (c)
\end{tabular}
\caption{(a),(b) Diagrams of group \grtwov{} and (c) group \grtwor{}
  contributing to the process $q \bar q \rightarrow \hz{}(g)$ at order
  $g^3\lambda_t\alpha_s^2$.}
\label{fig::zhaa}
\end{center}
\end{figure}

Examples of diagrams of the first group, named \gronev{} in what
follows, are shown in \fig{fig::vhneu}\,(a) and (b). They are
characterized by the emission of a Higgs boson off a top-quark
bubble-insertion into an {\it internal} (i.e.\ virtual) gluon line. They
contribute to the total cross section through the interference with the
leading order amplitude (see \fig{fig::logg}\,(a)).

The second group (\groner{}), see \fig{fig::vhneu}\,(c), can be viewed
as the real emission counterpart of group \gronev{}. It is obtained by
radiating the Higgs off a top-quark bubble-insertion into an {\it
external} gluon line.  These diagrams have to be interfered with the
real-emission amplitude contributing to the \nlo{} \qcd{} cross section,
see \fig{fig::logg}\,(b). Needless to say that the crossed amplitudes,
where the gluon is in- and a quark or anti-quark is out-going, have to
be taken into account as well.

The third group of Feynman diagrams (\grtwov{}) is closely related to the
gluon-induced contribution of \fig{fig::logg}\,(c). While the latter
enters the total cross section with its square, the new contributions
treated here are genuine two-loop terms which are to be interfered with
the tree-level amplitude of \fig{fig::logg}\,(a). Examples are shown in
\fig{fig::zhaa}\,(a) and (b).  Finally, the fourth group (\grtwor{}) can
again be seen as the real-emission counterpart of group \grtwov{}. An
example is shown in \fig{fig::zhaa}\,(c). Also here, of course, one can
cross the gluon and the quark or anti-quark from the final to the
initial state and vice versa.

The amplitude for each of these groups is separately gauge invariant and
{\abbrev UV}- as well as {\abbrev IR}-finite, despite the fact that two
of them can each be viewed as real and virtual correspondences of each
other.

The diagrams of \groner{} and \grtwor{} can be calculated exactly,
taking into account the full dependence on the top, Higgs, and vector
boson mass.  The amplitude arising from groups \gronev{} and \grtwov{},
on the other hand, is very challenging to compute and may be even beyond
current technology. We therefore follow the established and successful
method of asymptotic expansions in order to approximate the result in
the limit of infinite top-quark mass $\mtop{}$. The validity of this
approach will be discussed below.

\begin{figure}
\begin{center}
\begin{tabular}{ccc}
\includegraphics[width=.3\textwidth]{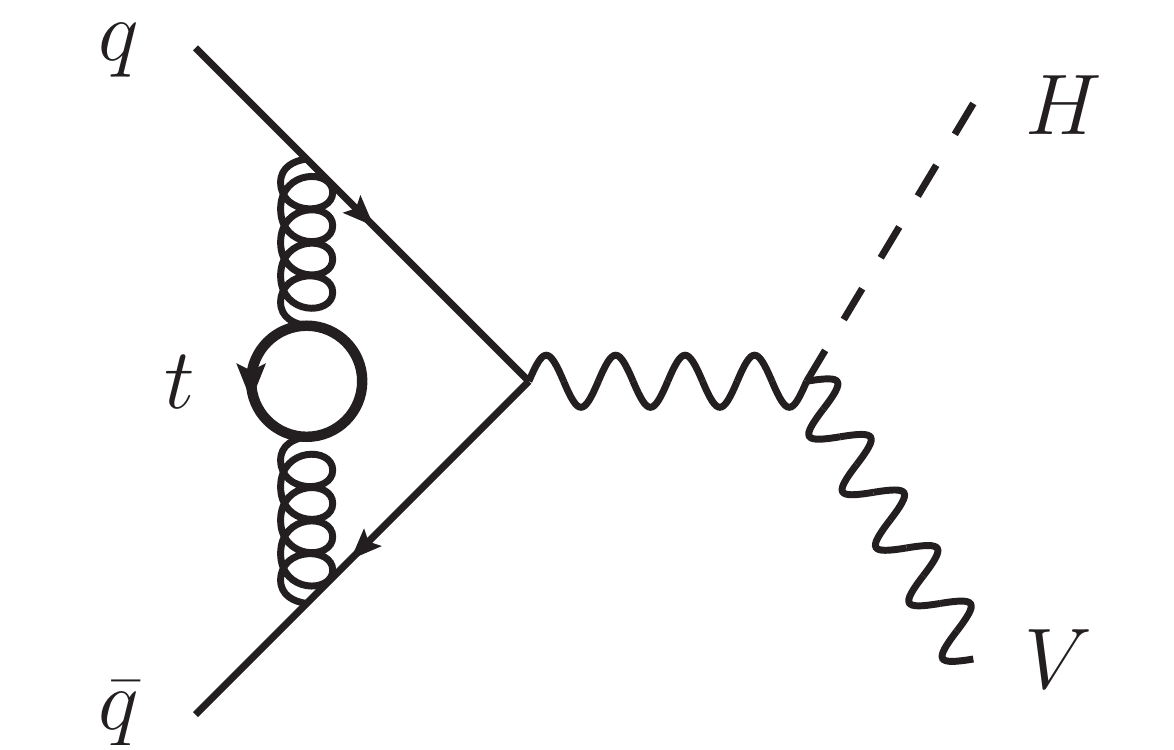} &
\includegraphics[width=.3\textwidth]{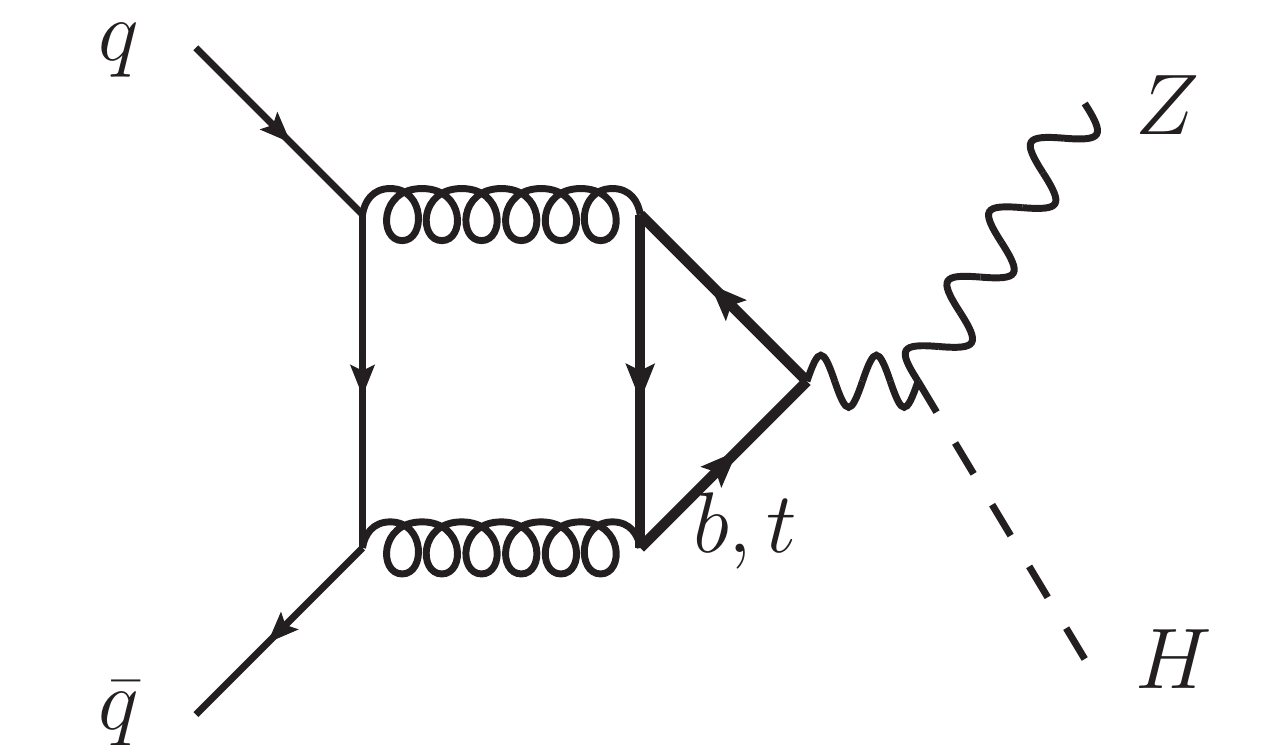} &
\raisebox{.5em}{\includegraphics[width=.3\textwidth]{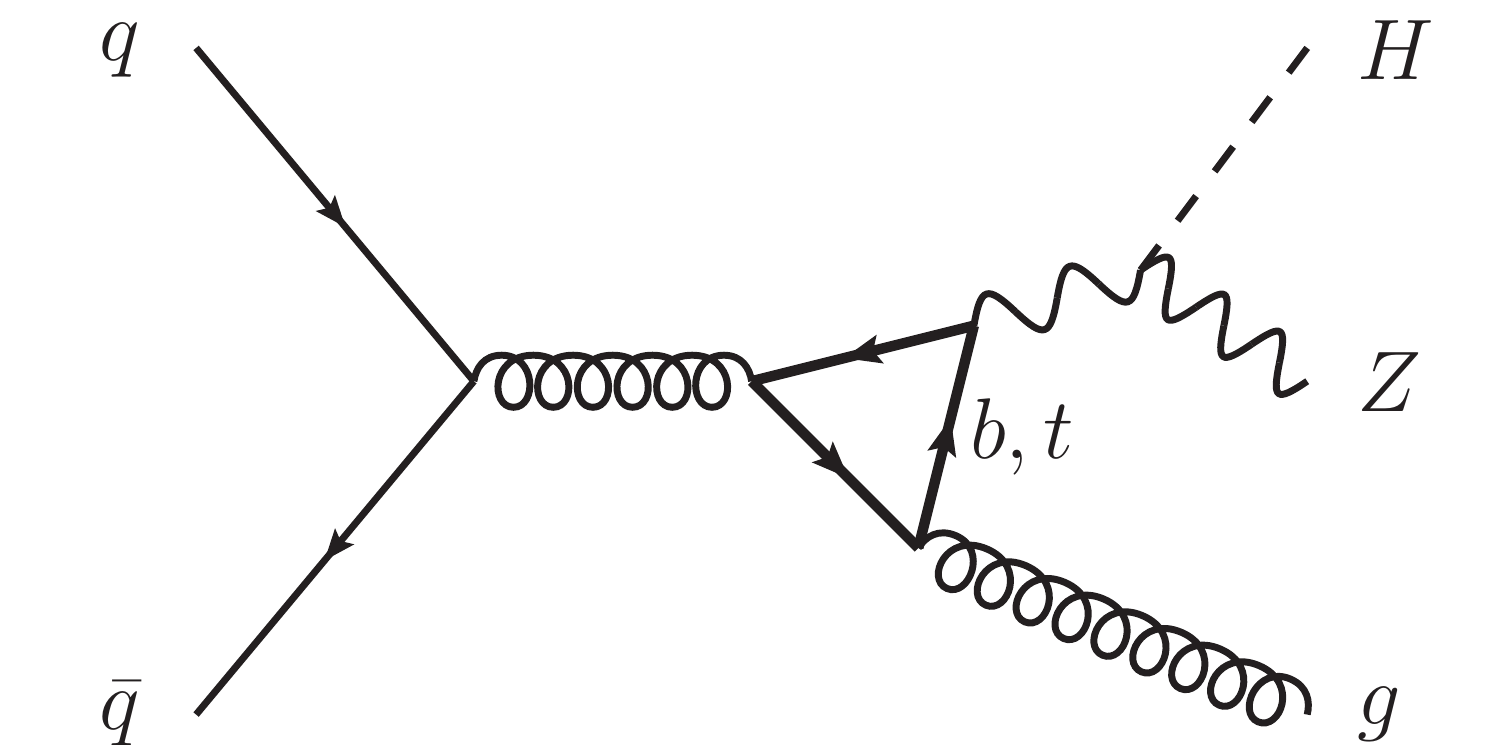}} \\
(a) & (b) & (c)
\end{tabular}
\caption[]{\label{fig::dytb}Drell-Yan-like diagrams with closed top- and
bottom-quark loops.}
\end{center}
\end{figure}

In a fully consistent treatment of the contributions described in this
section, one needs to take into account yet another set of Feynman
diagrams. They can be obtained from the previous ones by radiating the
Higgs boson off the external vector boson instead of the top-quark loop,
see \fig{fig::dytb}, for example.\footnote{Due to the requirement of
anomaly cancellation, however, they have to be supplemented by triangle
diagrams with bottom- instead of top-quarks running in the loop.}  If
all couplings are replaced by their \sm{} values, the resulting
amplitudes are of the same perturbative order as the ones discussed in
this paper; however, they receive an additional suppression factor $\sim
M_V^2/(\hat s-M_V^2)$, where $M_V$ is the mass of the vector boson and
$\sqrt{\hat s}\geq M_V+\mhiggs$ the partonic center-of-mass energy.
Justifiably so, these terms have been considered Drell-Yan-like in
Ref.~\cite{Brein:2003wg}; in fact, they can be calculated by convolving
the output of (a suitably modified version of) the program {\tt
MASSIVE}~\cite{Rijken:1995gi} with the decay $V^\ast\to VH$. As a check,
we also calculated some of these contributions directly using the
methods described below. The result confirms the statement of
Refs.~\cite{Brein:2003wg,Kniehl:1990iv} that these contributions are
numerically irrelevant: they are typically 2-3 orders of magnitude
smaller than the other newly evaluated contributions considered in this
paper, and will therefore be neglected in what follows.


\subsection{Calculation of \gronev{} and \grtwov{}}\label{sec::virtual}

\begin{figure}
\begin{center}
\begin{tabular}{cc}
\raisebox{-2.4em}{\includegraphics[width=0.14\textwidth]{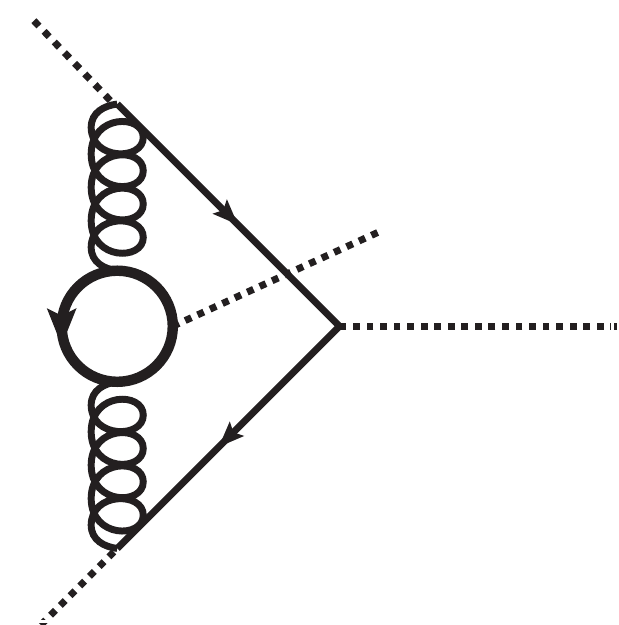}}
$\otimes$
\raisebox{-2.2em}{\includegraphics[width=0.2\textwidth]{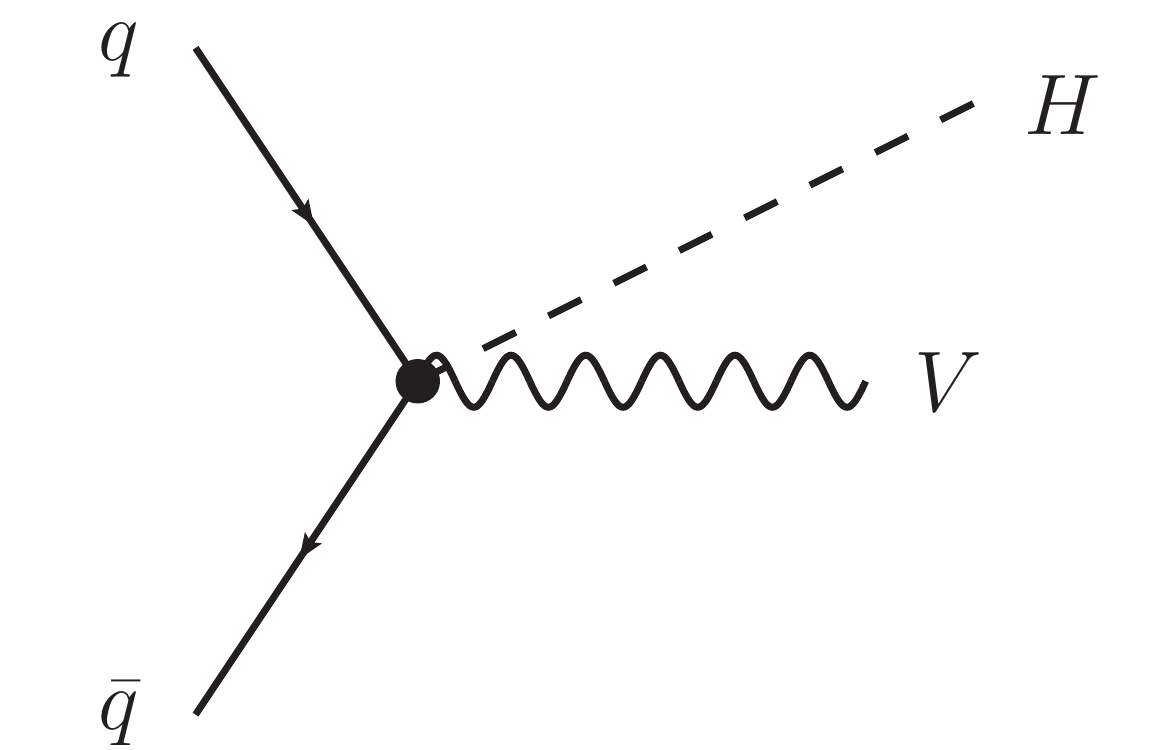}}
$+$ &
\raisebox{-1.3em}{\includegraphics[width=0.06\textwidth]{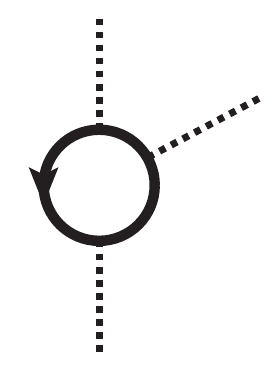}}
$\otimes$
\raisebox{-2.2em}{\includegraphics[width=0.2\textwidth]{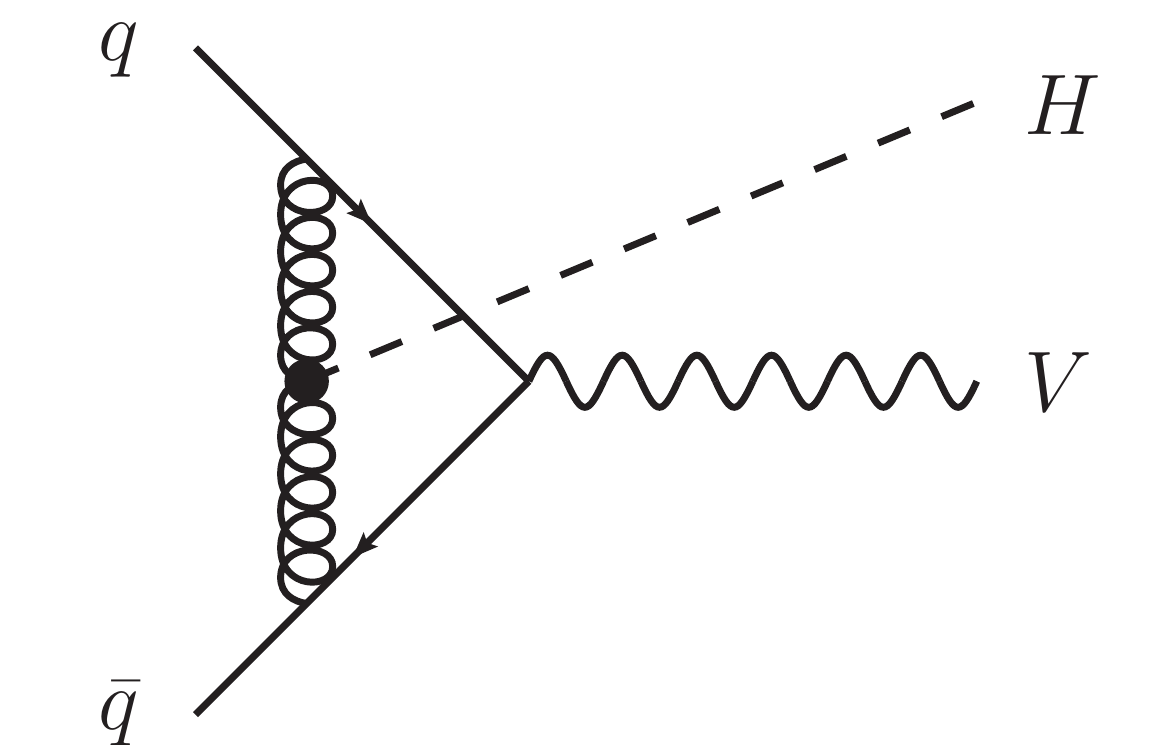}}
\end{tabular}
\caption{Asymptotic expansion of the diagram in
  \fig{fig::vhneu}\,(a). The diagrams left of $\otimes$ are evaluated
  after setting their external momenta to zero. The result determines
  the expression to be inserted into the effective $\bar qq\hv{}$ or
  $ggH$ vertex in the diagram right of $\otimes$. For details on the
  general method, see
  Refs.~\cite{Smirnov:1994tg,Smirnov:2002pj,Harlander:1999cs}, for example.}
\label{fig::asympt}
\end{center}
\end{figure}

A priori, an expansion of the two-loop amplitudes in the limit of large
top-quark mass seems unjustified, because the partonic center of mass
energy $\sqrt{\hat s}$ at the Tevatron and the \lhc{} can be much larger
than $\mtop$ (or rather $2\mtop$ which corresponds to the threshold and
is thus the relevant scale). Nevertheless, there is a number of
arguments that make such an approach reasonable, if $\mhiggs$ is not too
large:
\begin{itemize}
\item The parton luminosities at large $\hat s$ are strongly suppressed,
  and the bulk of the contribution to the total cross section indeed
  arises from the region below the top-quark threshold. This is similar
  to what happens in the case of gluon fusion, see
  Refs.~\cite{Harlander:2010my,Harlander:2009mq,Marzani:2008az}.
  For \higgsstrahlung{}, however, the situation is somewhat worse,
  because the energy window for which the heavy-top expansion is
  expected to converge is much narrower than for gluon fusion:
  $M_V+\mhiggs\leq\sqrt{\hat s}\leq 2\mtop$. In fact, for
  $\mhiggs\gtrsim 250$\,GeV, this restriction cannot be obeyed at all.
\item For group \gronev{} (\fig{fig::vhneu}\,(a),(b)), the typical energy scale
  affecting the top-quark loop is not the full center-of-mass energy,
  but significantly below that, because the vector boson carries off a
  large fraction of the momentum.
\item Since group \grtwov{} (\fig{fig::zhaa}\,(a),(b)) is closely
  related to the $gg$-induced contribution of \fig{fig::logg}\,(c), we
  may estimate its impact by the ratio of the $q\bar q$ and the $gg$
  luminosity times the $gg$-induced cross section. It should therefore
  not exceed a few percent of the \lo{} cross section.
\item Overall, we expect the top-mediated terms to affect the cross
  section at the percent level~\cite{Ferrera:2011bk,Hirschi:2011pa}.
  The leading term of an expansion in terms of $1/\mtop{}$ should thus
  be sufficiently precise.
\end{itemize}
The tools to calculate the diagrams are by now standard: asymptotic
expansion of the two-loop diagrams of groups \gronev{} and \grtwov{}
leads to a factorization of scales into either two-loop massive tadpoles
(i.e., vanishing external momenta) times tree-level diagrams with an
effective $\bar qqH$ or $\bar qq\hv{}$ vertex, or one-loop massive
tadpoles times massless one-loop diagrams with an effective $gg\hz{}$ or
$ggH$ vertex.  As an example, the graphical representation of the
asymptotic expansion of the diagram in \fig{fig::vhneu}\,(a) is shown
in \fig{fig::asympt}.

We use the automatic setup consisting of {\tt
  qgraf}~\cite{Nogueira:1993ex} for the generation, and {\tt
  q2e/exp}~\cite{Harlander:1997zb,Seidensticker:1999bb} for the
expansion of the diagrams, as well as {\tt
  MATAD}~\cite{Steinhauser:2000ry} for the calculation of the tadpole
integrals. In order to evaluate the massless one-loop box and triangle
integrals corresponding to the right-most diagrams in \fig{fig::asympt},
we supplement this setup by an additional routine (based on {\tt
  FORM}~\cite{Vermaseren:2000nd}) which implements standard
Passarino-Veltman reduction~\cite{Passarino:1978jh} in algebraic
form. The scalar one-loop Feynman integrals are evaluated using the
results of Ref.~\cite{Ellis:2007qk}.

It turns out that for \gronev{}, the sum of terms involving an effective
$\bar qq\hv{}$ or $\bar qqH$ vertex does not contribute at leading order
in $1/\mtop{}$, i.e., the corrections can be calculated by simply using
an effective $ggH$ vertex and evaluating one-loop diagrams. This
observation allowed us to obtain \gronev{} in a second, independent
calculation by using a generalized version of the {\tt FORM} program
{\tt FDiag}~\cite{FDiag} and the Fortran package {\tt
FF}~\cite{vanOldenborgh:1989wn,FF,Aeppli}, supplemented by a routine to
facilitate the tensor reduction of rank-4 tensor 4-point functions and
the subsequent numerical evaluation of the corresponding tensor
coefficients.

Contracting with the \lo{} amplitude and summing/averaging over
final/initial color and spin degrees of freedom, we find for the
amplitude of \gronev{}, cf.\,\fig{fig::vhneu}\,(a),(b):
\begin{equation}
\begin{split}
-\frac{\dd \Delta\hat\sigma^V_\text{I}}{\dd \hat t} &=
G_{qq'}^V\frac{\gfermi^2 M_V^4}{108\pi\hat s^2(\hat
  s-M_V^2)}\left(\api\right)^2 \,\Re \Biggl\{ -2\hat s \\&+\ln \left(
\frac {\hat u} {M_V^2} \right) \left(\hat t-M_V^2 + \frac{2 \hat s
  M_V^2}{\hat u - M_V^2} \right) +\ln \left( \frac {\hat t} {M_V^2}
\right) \left(\hat u -M_V^2 + \frac{2 \hat s M_V^2}{\hat t - M_V^2}
\right) \\ & + \left( \hat s - \mhiggs^2 + M_V^2 \right) \left[
  \ln^2\left(\frac {\hat t}{\hat u} \right) + 2 \dilog \left(1 - \frac
     {\mhiggs^2} {\hat t}\right) + 2 \dilog \left(1 - \frac {\mhiggs^2}
     {\hat u}\right) \right.  \\ & \left.  + 2\dilog \left(1 - \frac
     {M_V^2} {\hat t}\right) + 2 \dilog\left(1 - \frac {M_V^2} {\hat
       u}\right) - 2 \dilog\left(1 - \frac {\mhiggs^2 M_V^2} {\hat t
       \hat u}\right) \right] + \order{\frac{1}{\mtop^2}}\Biggr\}\,,
\label{eq::ds1dt}
\end{split}
\end{equation}
where $\hat s=(p_1+p_2)^2$, $\hat t=(p_1-p_V)^2$ and $\hat
u=(p_1-p_H)^2$ are the usual (partonic) Mandelstam variables, with the
incoming momenta $p_1,p_2$, and $p_H,p_V$ the momenta of the Higgs and
the vector boson, respectively.  $M_V$ is the mass of
the emitted vector boson, and $\dilog$ denotes the di-logarithm. The
electro-weak couplings are given by
\begin{equation}
\begin{split}
G_{qq'}^Z &= (v_q^2 + a_q^2)\delta_{qq'}\,,
\qquad G_{qq'}^W = \frac{1}{2}|V_{qq'}|^2
\,,\\
v_q &= \pm\frac{1}{2} + \frac{1}{3}\left\{\begin{array}{c}
-4\\+2\end{array}\right\}\sin^2\theta_W\,,\quad
a_q = \pm\frac{1}{2}
\qquad
\text{for}\quad q\in\left\{
\begin{array}{c}
u,c\\
d,s,b
\end{array}
\right\}\,,
\end{split}
\end{equation}
with the weak mixing angle $\sin^2\theta_W=1-M_W^2/M_Z^2$ and the
{\abbrev CKM} matrix elements $V_{qq'}$ (see
Ref.~\cite{Nakamura:2010zzi} for the latest numerical values; we set
$V_{qq'}=0$ if both $q$ and $q'$ carry the same weak isospin charge $I_3$).
We have expressed the top-quark Yukawa coupling by the tree-level relation
\begin{equation}
\begin{split}
\lambda_t\equiv \frac{\mtop}{v} = \mtop\sqrt{\sqrt{2}\gfermi}
\end{split}
\end{equation}
in \eqn{eq::ds1dt}, and also in \eqn{eq::ds3dt} below. Similar to the
gluon fusion process, the factor $\mtop$ cancels against its inverse
from the top-loop integration.

For group \grtwov{} (see \fig{fig::zhaa}\,(a),(b)), only the axial
vector part of the $Z$ coupling to fermions contributes. We implement
its Dirac structure with the help of Levi-Civita symbols
$\varepsilon_{\mu\nu\rho\sigma}$~\cite{Larin:1993tq}
\begin{equation}
\begin{split}
\gamma_\mu\gamma_5 \to \frac{i}{3!}\varepsilon_{\mu\alpha\beta\delta}
\gamma^\alpha\gamma^\beta\gamma^\delta\,,
\end{split}
\end{equation}
and re-write their product in terms of the anti-symmetrized (denoted by
square brackets) $D$-dimensional metric tensor,
\begin{equation}
\begin{split}
\varepsilon_{\mu\nu\rho\sigma}
\varepsilon^{\alpha\beta\gamma\delta} = 
-g^{[\alpha}_{[\mu}
g^{\beta\phantom{[}}_{\nu\phantom{]}}
g^{\gamma\phantom{[}}_{\rho\phantom{]}}
g^{\delta]}_{\sigma]}\,.
\end{split}
\end{equation}

Upon asymptotic expansion of the diagrams, the terms corresponding to an
effective $gg\hz{}$ vertex contribute only at subleading order in
$1/\mtop$, and the calculation reduces to massive 2-loop tadpole
diagrams.

The result assumes the simple form
\begin{equation}
\begin{split}
-\frac{\dd \Delta\hat\sigma^Z_\text{II}}{\dd \hat t} &=
\frac{\gfermi^2 M_Z^4 a_t a_q}{12\pi\hat s^2(\hat
  s-M_Z^2)}\left(\api\right)^2 \Bigg\{2 \hat s - \mhiggs^2 +\frac{\hat t
  \hat u}{M_Z^2} + \order{\frac{1}{\mtop}}\Bigg\}\,.
\label{eq::ds3dt}
\end{split}
\end{equation}

Since the expressions in \eqn{eq::ds1dt} and (\ref{eq::ds3dt}) are free
of divergences in the allowed $\hat t$-region, we can numerically
integrate them together with the convolutions over the \pdf{}s in order
to get their contribution to the total inclusive hadronic cross
section. The numerical results will be presented in the next section.


\subsection{Calculation of \groner{} and \grtwor{}}\label{sec::real}

Being of one-loop order, the diagrams of groups \groner{} and \grtwor{}
can be calculated including the full dependence on the top-quark,
Higgs- and vector-boson mass by means of Passarino-Veltman reduction to
scalar one-loop functions.  In most of the phase-space, the numerical
evaluation of these function can be performed with the help of the
programs {\tt FDiag} and {\tt FF} (see Sec.\,\ref{sec::virtual}), and
independently with {\tt FeynArts}, {\tt FormCalc}, and {\tt
  LoopTools}~\cite{feynarts} (the latter of which relies on {\tt FF}
though). The contribution \grtwor{}, however, involves momentum
configurations that are outside {\tt FF}'s capabilities. For those, we
use an implementation of the one-loop integrals based
on~Refs.~\cite{Denner:2005nn,Denner:2010tr}\footnote{We thank Stefan
  Dittmaier for providing us with his private code.}. Again, since
projection with the \lo{} amplitude leads to a finite expression,
phase-space integration and convolution with \pdf{}s can be done fully
numerically.


\subsection{Validation of the heavy-top approximation}\label{sec::validation}

This section describes both a consistency check for our calculation of
the real-emission contributions \groner{} and \grtwor{}, as well as a validity
check of our approach to the virtual terms \gronev{} and \grtwov{}.

In addition to the exact calculation as described in
Section~\ref{sec::real}, we evaluated the real emission amplitudes
\groner{} and \grtwor{} also by applying asymptotic expansions in the
heavy-top limit.  In these cases, it reduces to a naive Taylor expansion
of the integrand before loop integration.  We can again perform all
phase-space integrals numerically. In the $q\bar q$-channel, however, we
also integrated over the angular variables of the phase-space
analytically in $D=4-2\ep$ space-time dimensions, and found the
cancellation of all poles at $\ep=0$.  The numerical integration over
the remaining energy variables is straightforward and leads to the same
result as the all-numerical method.

Comparing this heavy-top result for \groner{} and \grtwor{} at the
partonic level to the exact results described above, we indeed observe
that heavy-top limit systematically approaches the exact result as
${\cal E}/\mtop\to 0$, where ${\cal E}$ is any of the external energy
scales. In particular, we observe that, as opposed to the other
contributions, the terms \grtwor{} vanish as $1/\mtop^2$ in the limit
$\mtop\to \infty$. This agreement between the exact and the asymptotic
approach provides a strong and valuable check on our results.

In order to draw conclusions for the validity of the heavy-top
expansion, however, the relevant quantity to compare is the hadronic
cross section, of course. For that, we find that the heavy-top result is
roughly within 25\%/35\% of the full result at the \lhc{}/Tevatron in
the mass range considered in this paper.

We expect a similar quality of the heavy-top result for the
contributions from \gronev{} and \grtwov{}. To be conservative, we
attribute an additional uncertainty of 30\%/50\% (relative to the
central values) to these terms for the \lhc{}/Tevatron prediction of the
hadronic cross section.



\section{Numerics}\label{sec::numerics}


\subsection{Size of the top-induced terms}\label{sec::topnum}

We present numbers for the Tevatron ($p\bar p$ @ $\sqrt{s} =
1.96$\,TeV), as well as the \lhc{} ($pp$ @ 7 \,TeV and 14\,TeV).  The
hadronic cross section is evaluated by folding the partonic cross
section with \pdf{}s. Although the corrections evaluated in this paper
are not renormalized by lower order terms, we think it is appropriate to
convolve them with \nnlo{} \pdf{}s. We use the central {\abbrev
  MSTW2008} set, implying $\alpha_s(M_Z) = 0.1171$.  For the physical
parameters, we assume the values
\begin{equation}
\begin{split}
M_Z = 91.1876\,\text{GeV}\,,\qquad M_W = 80.398\,\text{GeV}\,.
\end{split}
\end{equation}
Since the amplitudes considered in this paper are {\abbrev UV}- and
{\abbrev IR}-finite, they do not depend explicitly on the
renormalization or factorization scale ($\muR$, $\muF$). However, these
scales enter implicitly through the strong coupling and the
\pdf{}s. For the numerical analysis, we choose
\[
q^2\equiv(p_{H}+p_{V})^2
\]
as the central scale for $\muR^2$ and $\muF^2$, where $p_{H}$ and
$p_{V}$ are the 4-momenta of the Higgs and the outgoing vector
boson. For the Drell-Yan-like terms, $\sqrt{q^2}$ equals the invariant
mass of the intermediate gauge boson. In order to estimate the
theoretical uncertainty of our results, we vary $\muR$ by a factor of
three around this central scale while keeping $\muF$ fixed,
\begin{equation}
\begin{split}
\frac{1}{3}\sqrt{q^2}\leq
\muR\leq 3\sqrt{q^2}\,,\qquad \muF=\sqrt{q^2}\,,
\label{eq::scale}
\end{split}
\end{equation}
then repeat the analysis after interchanging $\muF$ and $\muR$, and take
the extreme values of the cross section as uncertainty band.

\begin{figure}
  \begin{center}
    \begin{tabular}{cc}
      \includegraphics*[width=.5\textwidth,
        viewport=0 0 290 250]{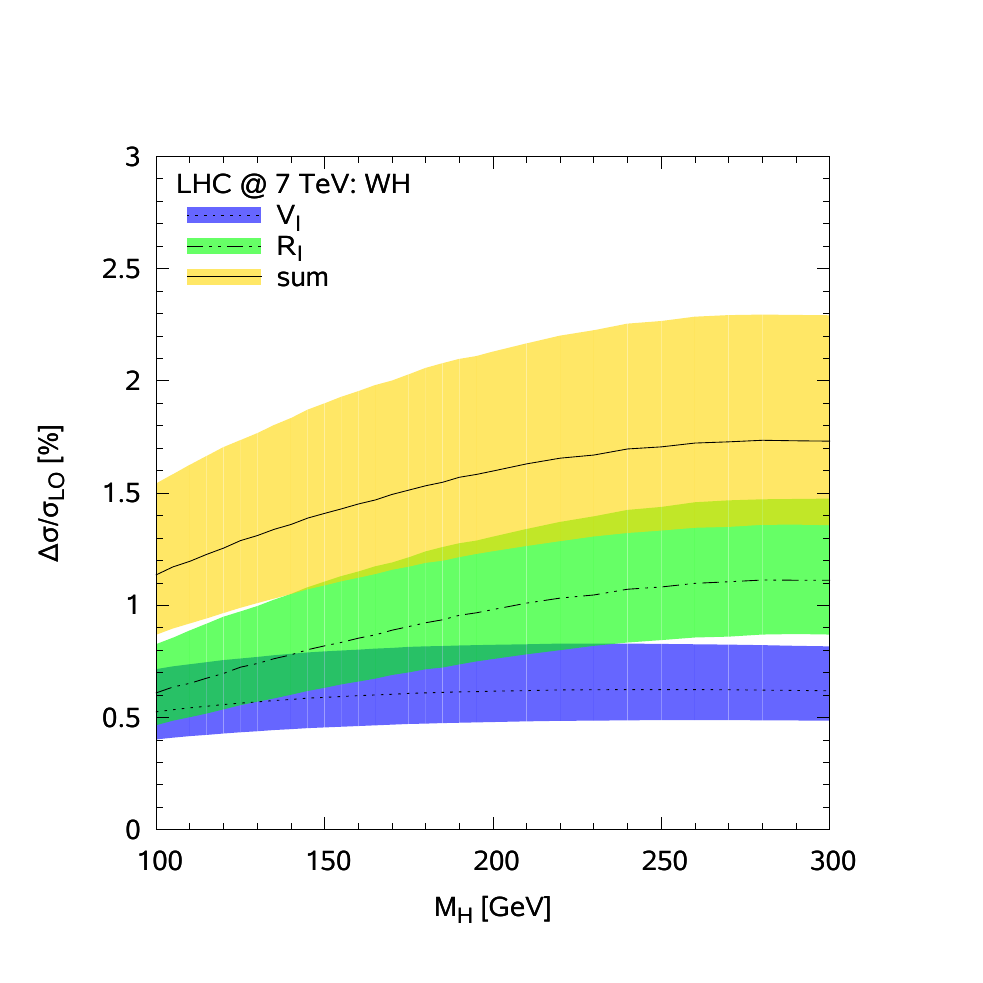} &
      \includegraphics*[width=.5\textwidth,
        viewport=0 0 290 250]{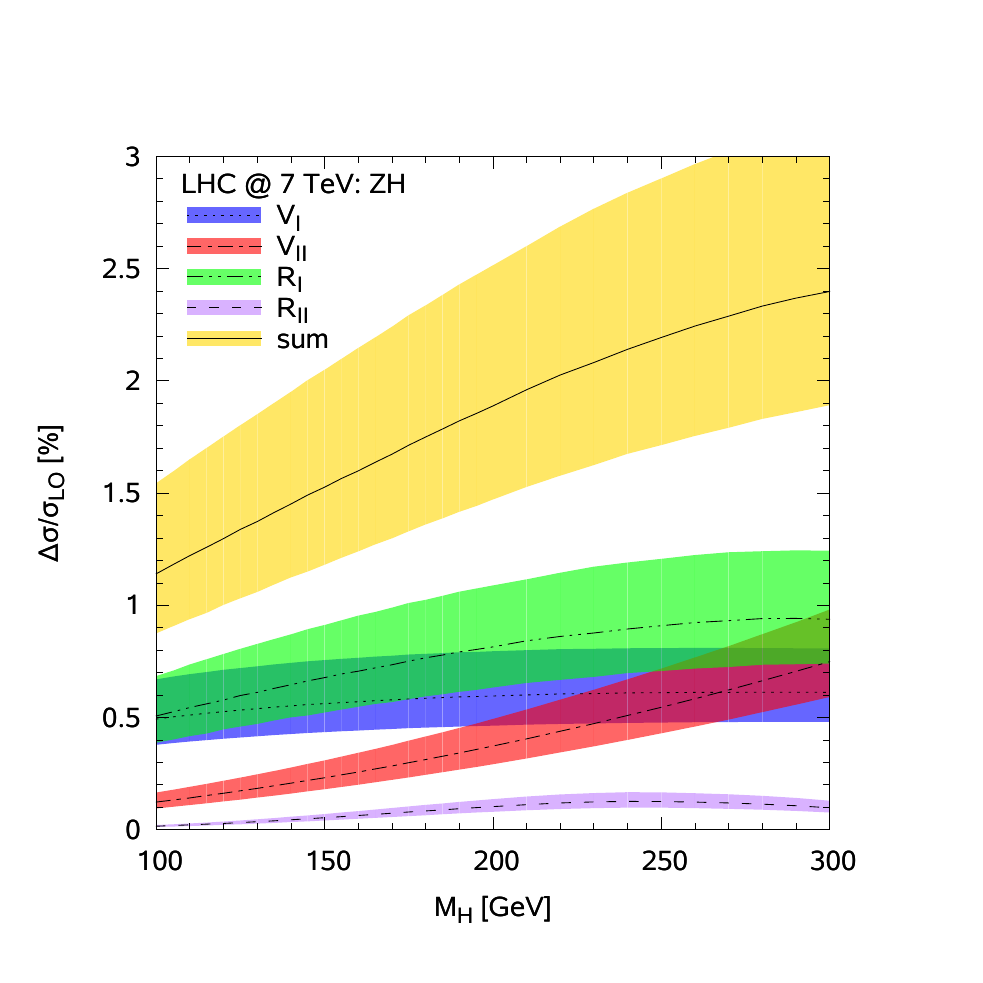} \\[-1em]
      (a) & (b) \\[.5em]
      \includegraphics*[width=.5\textwidth,
        viewport=0 0 290 250]{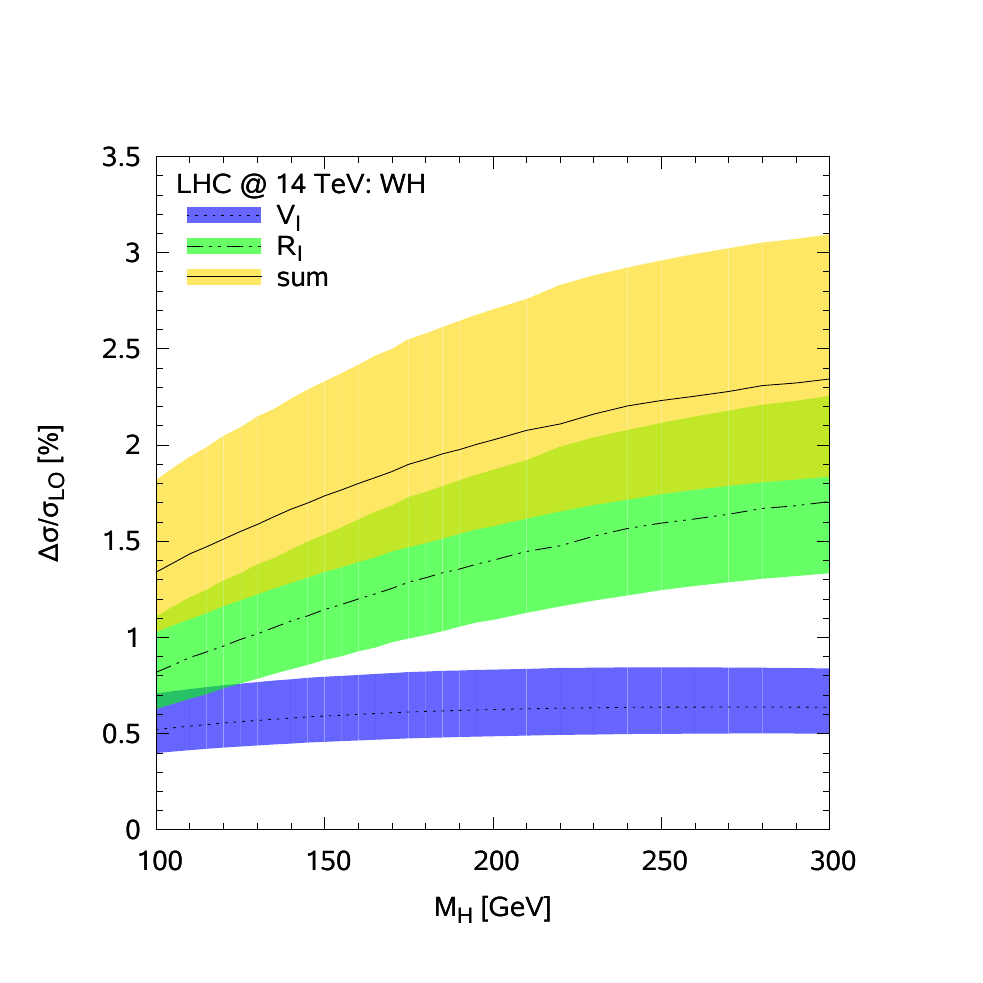} &
      \includegraphics*[width=.5\textwidth,
        viewport=0 0 290 250]{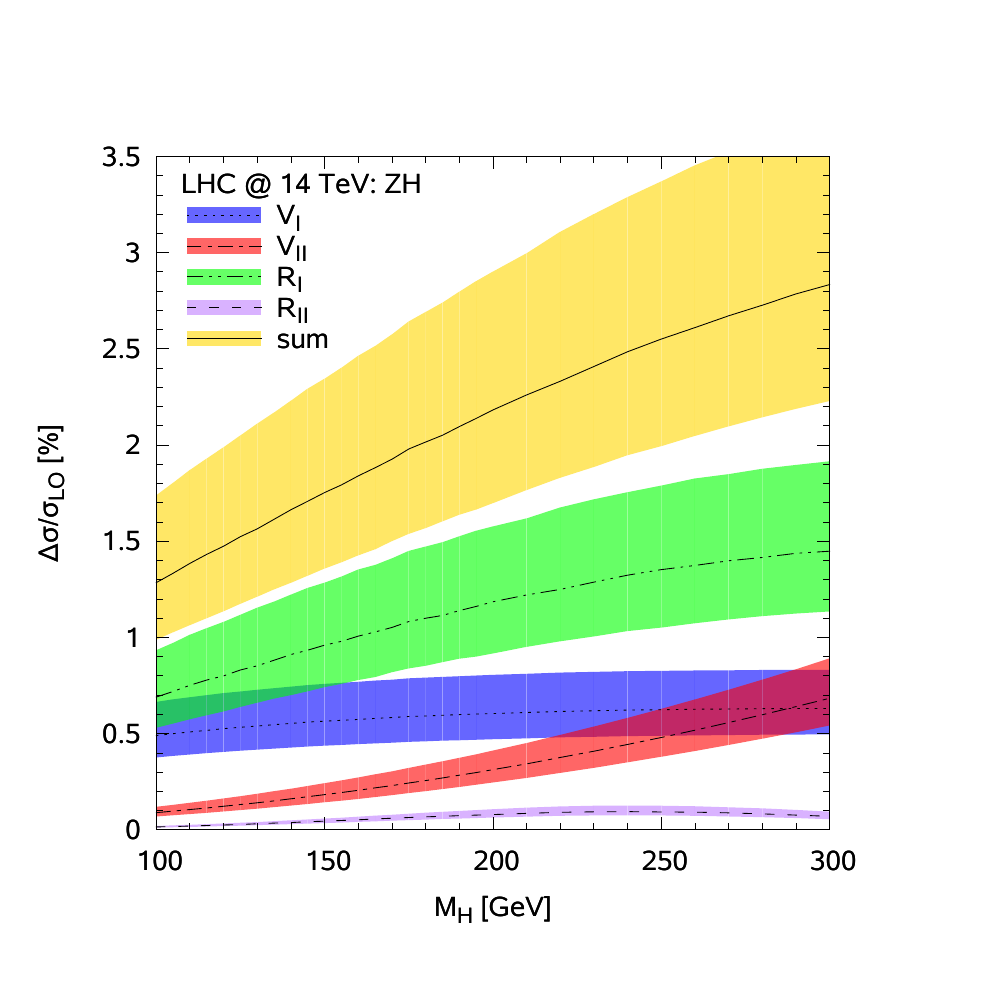} \\[-1em]
      (c) & (d)
    \end{tabular}
    \parbox{.9\textwidth}{
      \caption[]{\label{fig::sigrellhc}\sloppy Contribution of the
        corrections evaluated in this work to the total inclusive cross
        section $\sigma(pp\to \hw{}+X)$ (left column) and $\sigma(pp\to
        \hz{}+X)$ (right column) at the \lhc{} with 7\,TeV (upper row)
        and 14\,TeV center-of-mass energy (lower row). The effects are
        shown relative to the leading order cross section
        $\sigma_\text{\lo{}}$. The various bands show the contributions
        from \gronev{}, \groner{}, and, in the case of $\hz{}$,
        \grtwov{} and \grtwor{}. The upper band is the sum of all
        contributions. The width of the band arises from the variation
        of $\muR$ and $\muF$ as described in the main text.}}
  \end{center}
\end{figure}

\begin{figure}
  \begin{center}
    \begin{tabular}{cc}
      \includegraphics*[width=.5\textwidth,
        viewport=0 0 290 250]{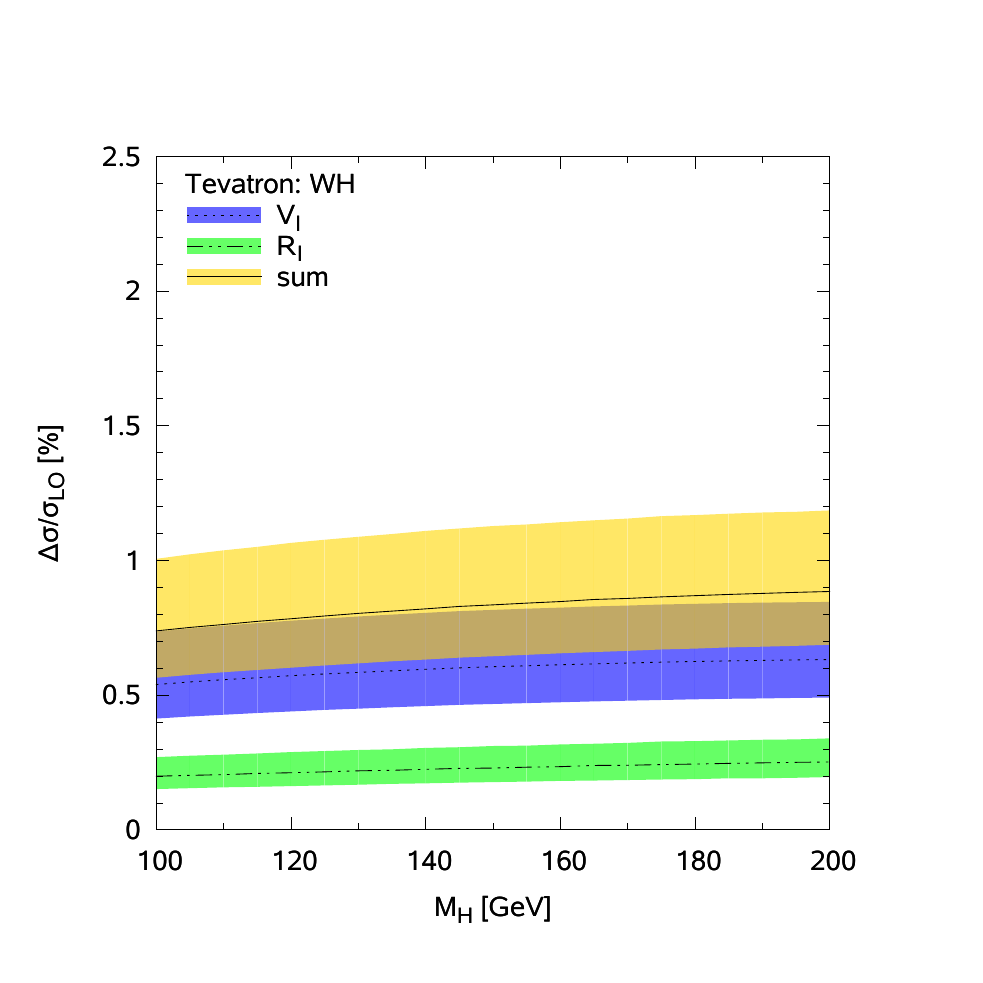} &
      \includegraphics*[width=.5\textwidth,
        viewport=0 0 290 250]{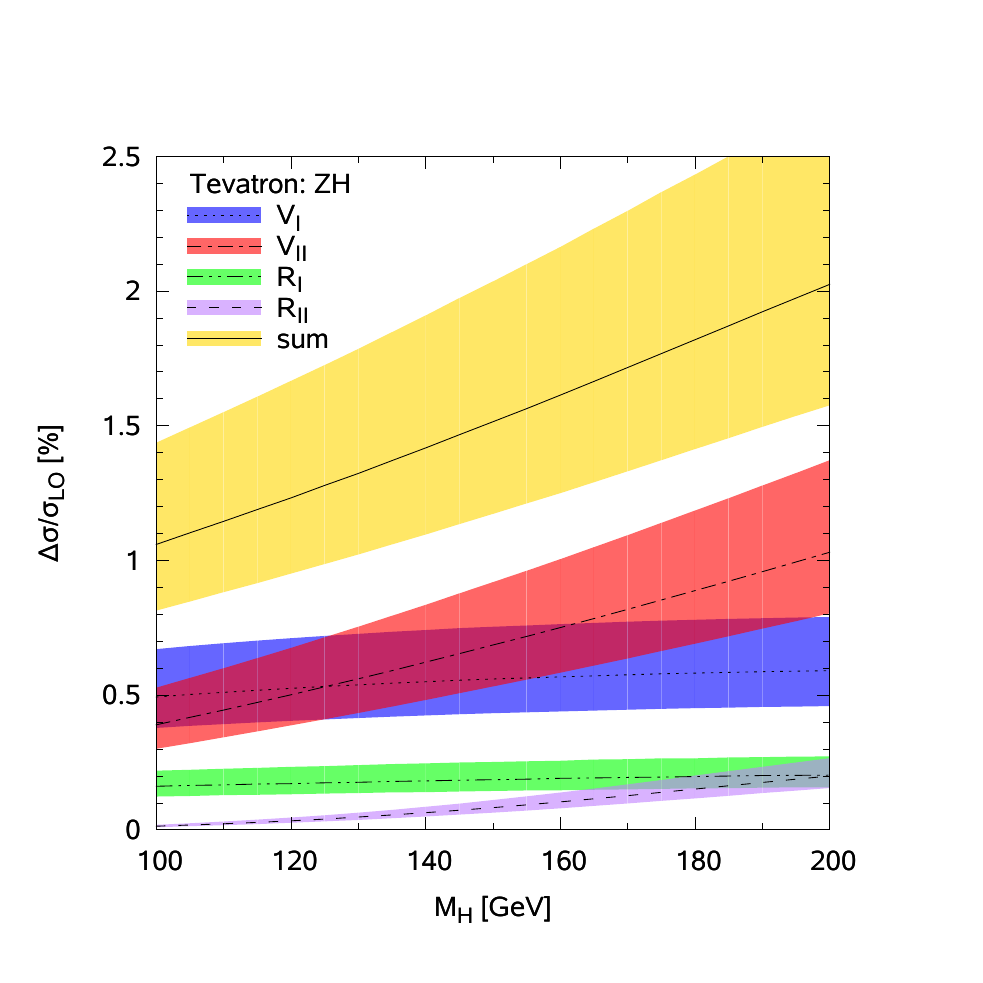} \\[-1em]
      (a) & (b)
    \end{tabular}
    \parbox{.9\textwidth}{
      \caption[]{\label{fig::sigreltev}\sloppy Same as
        \fig{fig::sigrellhc}, but for $\hw{}$/$\hz{}$ production at the
        Tevatron.}}
  \end{center}
\end{figure}

\fig{fig::sigrellhc} shows the contribution of the newly evaluated terms
to the total inclusive cross sections $\sigma(pp\to \hw{})\equiv\sigma(pp\to
W^+H)+\sigma(pp\to W^-H)$ and $\sigma(pp\to \hz{})$ at 7\,TeV and
14\,TeV center-of-mass energy. The contributions from the various groups
of diagrams (see Section~\ref{sec::outline}) are shown separately, each
of them with an error band derived from the scale variation described
above. Also shown is the sum of all contributions. The corresponding
plots for the Tevatron are shown in \fig{fig::sigreltev}.

The size of the \gronev{} component amounts to about 0.5\% of the
\lo{} cross section\footnote{Only central values are considered in this
  discussion.} both for $\hw{}$ and $\hz{}$ production, independent of
collider type, center-of-mass energy, and Higgs boson mass. While at the
\lhc{} the \groner{} terms are typically a little larger than that, and
increase with $\mhiggs$ and the center-of-mass energy, they always
remain below 0.5\% at the Tevatron.

Since \gronev{} and \groner{} are the only non-vanishing contributions
for $\hw{}$ production, in this case the overall effect of the
top-mediated terms evaluated in this paper remains below 1\% at the
Tevatron, and ranges between about 1.1\% (for $\mhiggs=100$\,GeV at
7\,TeV) to 2.4\% (for $\mhiggs=300$\,GeV at 14\,TeV) at the \lhc{} (see
\fig{fig::sigrellhc}\,(a) and \fig{fig::sigreltev}\,(a)).

For $\hz{}$ production, also the groups \grtwov{} and \grtwor{} have to
be taken into account. Being suppressed by $1/\mtop^2$ at large $\mtop$,
as pointed out above, the terms \grtwor{} have only a very small
numerical impact at the per-mille level. The relative contribution of
\grtwov{}, on the other hand, increases with the Higgs boson mass, but
remains below 0.8\% at the \lhc{} in all of the considered Higgs mass
range. At the Tevatron, however, it exceeds \gronev{} and \groner{}
above $\mhiggs\approx 130$\,GeV and ranges up to about 1\% at
$\mhiggs=200$\,GeV. The overall effect of \gronev{}, \groner{},
\grtwov{}, and \grtwor{} on $\hz{}$ production at the Tevatron is
therefore between 1\% and 2\%, while it ranges between 1.1\% (for
$\mhiggs=100$\,GeV at 7\,TeV) and 2.9\% (for $\mhiggs=300$\,GeV at
14\,TeV) at the \lhc{} (see
\fig{fig::sigrellhc}\,(b) and \fig{fig::sigreltev}\,(b)).

In each case, a rough estimate of the uncertainty on the top-mediated
terms due to scale variation is 20-30\%. As discussed in
Section~\ref{sec::validation}, we also include an estimated uncertainty
on the \gronev{} and \grtwov{} terms of 30\%/50\% for the
\lhc{}/Tevatron, arising from the heavy-top limit (this is not included
in Figs.\,\ref{fig::sigrellhc} and \ref{fig::sigreltev}). Considering
the fact that the corrections are at the percent level, this uncertainty
will affect the accuracy of the total inclusive cross section by roughly
0.5\%. We do not expect the \pdf{} uncertainties on the top-induced
terms to add significantly to the one of the total cross section and
will neglect it in what follows.


%
%


%
%



\subsection{Total inclusive cross section}

As shown in the previous section, the effects of the newly evaluated
terms are of the order of 1-3\% of the \lo{} cross section on $\hw$
and $\hz$ production. Superficially, this lies within previous estimates
of the uncertainties of the total inclusive cross section at the
\lhc{}~\cite{Dittmaier:2011ti}. However, most of this uncertainty is
induced by the \pdf{}s. Once systematical effects and theoretical
ambiguities will have been settled (for a recent discussion, see
Ref.~\cite{Thorne:2011kq}) or become irrelevant due to the inclusion of
\lhc{} data, one may expect these uncertainties to shrink
significantly. Then, in order to arrive at the most precise prediction for the
process, the effects of the top-mediated terms evaluated here will have
to be included.

For this reason, we provide an updated prediction of the total inclusive
cross section for $\hw{}$ and $\hz$ production at the Tevatron and the
\lhc{} at 7\,TeV and 14\,TeV in this section. The numbers for the \lhc{}
are obtained simply by adding the top-mediated terms to the predictions
of Ref.~\cite{Dittmaier:2011ti}, which have been obtained using the
program {\tt vh@nnlo}~\cite{vhnnlo}, based on the calculation of
Refs.~\cite{Brein:2003wg,Hamberg:1991np,Harlander:2002wh} and the
program {\tt zwprod}~\cite{Hamberg:1991np}, and an electro-weak
correction factor~\cite{Ciccolini:2003jy}. The uncertainties on the
numbers given here result from linearly adding the previous
uncertainties~\cite{Dittmaier:2011ti} to the ones discussed in
Section~\ref{sec::topnum}.

For the Tevatron, an ``official'' public agreement on the cross section
prediction analogous to Ref.~\cite{Dittmaier:2011ti} is not
available. We therefore follow the analysis of
Ref.~\cite{Dittmaier:2011ti} using Tevatron parameters, and proceed as
above. The electro-weak correction factor we read off from the plots and
tables of Ref.~\cite{Ciccolini:2003jy} in this case, despite the fact
that the parameters in this paper are slightly outdated. The uncertainty
induced by this procedure should be negligible, however.

The final results are presented in Tables~\ref{tab::lhc7},
\ref{tab::lhc14}, and \ref{tab::tevatron} for the \lhc{} at 7\,TeV,
14\,TeV, and the Tevatron, respectively. The top-mediated terms slightly
increase all the cross sections, but also the theoretical
uncertainty. The column labeled ``Pert'' in the tables contains the
estimate of the perturbative uncertainty as described in
Section~\ref{sec::topnum}. The column labeled ``\pdf{}+$\alpha_s$''
displays the error induced by the \pdf{}s and the input value for
$\alpha_s$, as described in Ref.~\cite{Dittmaier:2011ti}.

\newcommand{\MH}{\mhiggs}
\begin{table}
\begin{center}
\include{inputs/wzh7nnlo-fb}
\end{center}
\caption[]{\label{tab::lhc7}Numerical values for the total cross section
  at \lhc{} at 7\,TeV.}
\end{table}

\begin{table}
\begin{center}
\include{inputs/wzh14nnlo-fb}
\end{center}
\caption[]{\label{tab::lhc14}Numerical values for the total cross section
  at \lhc{} at 14\,TeV.}
\end{table}

\begin{table}
\begin{center}
\include{inputs/wzhtevatronnnlo-fb}
\end{center}
\caption[]{\label{tab::tevatron}Numerical values for the total cross section
  at the Tevatron.}
\end{table}

\begin{figure}
  \begin{center}
    \begin{tabular}{cc}
      \includegraphics[width=.45\textwidth]{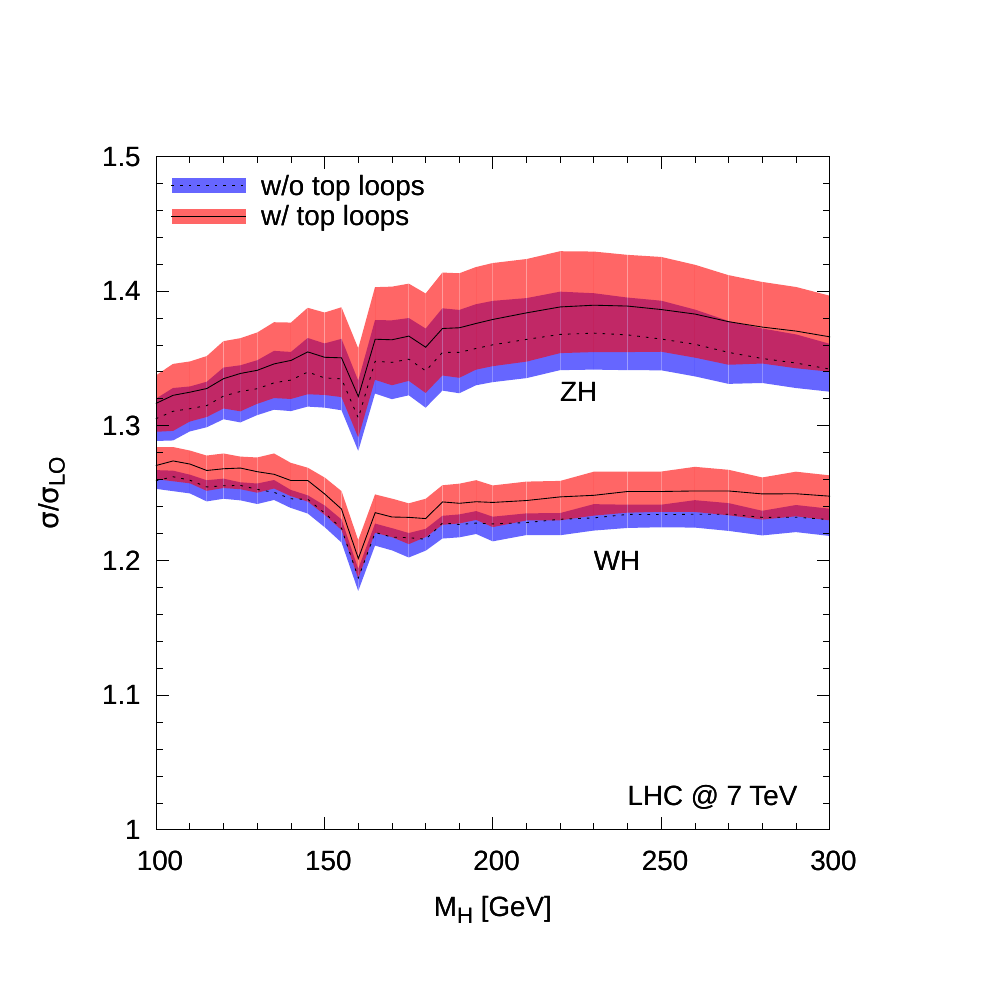} &
      \includegraphics[width=.45\textwidth]{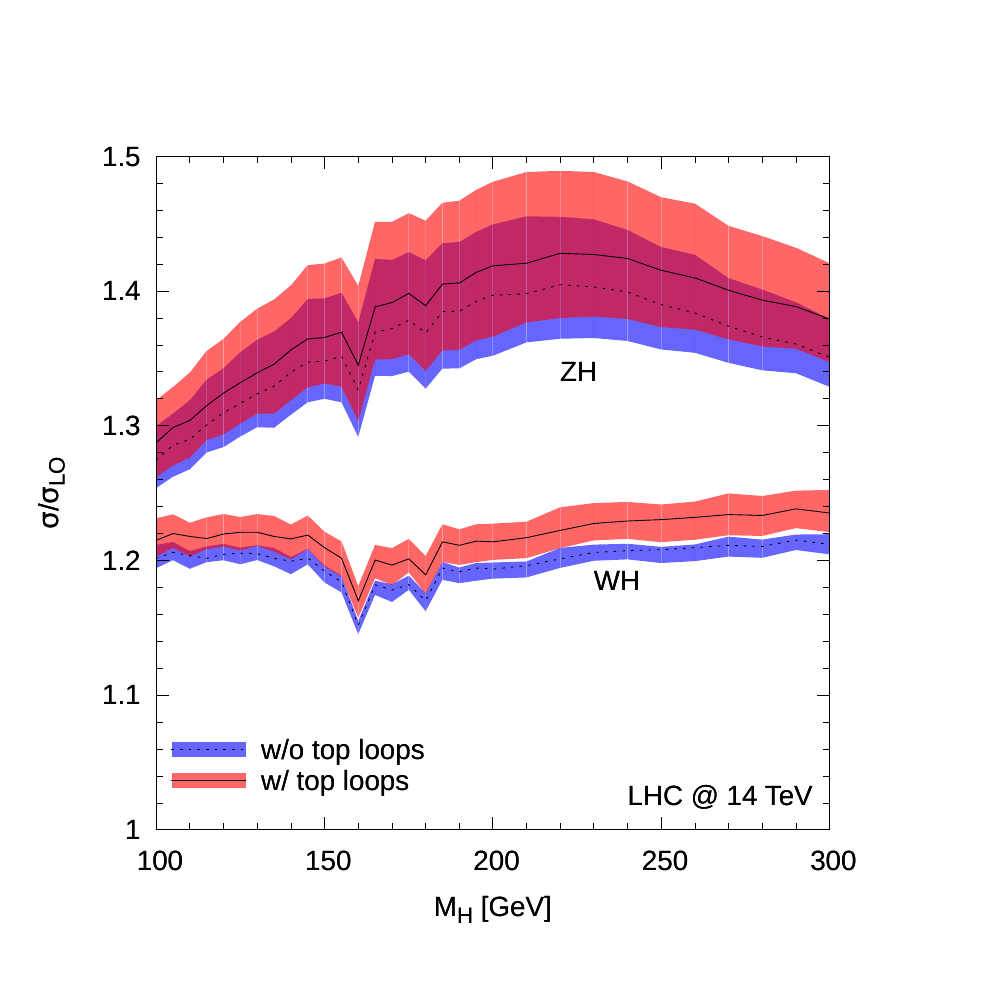}
      \\      (a) & (b)
    \end{tabular}
    \parbox{.9\textwidth}{
      \caption[]{\label{fig::kfaclhc}\sloppy K-factor at the \lhc{} at
        (a) 7\,TeV and (b) 14\,TeV center-of-mass energy, including
        \nnlo{} \qcd{} and electro-weak corrections, with and without
        the newly evaluated top-quark induced terms.  }}
  \end{center}
\end{figure}

\begin{figure}
  \begin{center}
    \begin{tabular}{cc}
      \includegraphics[width=.45\textwidth]{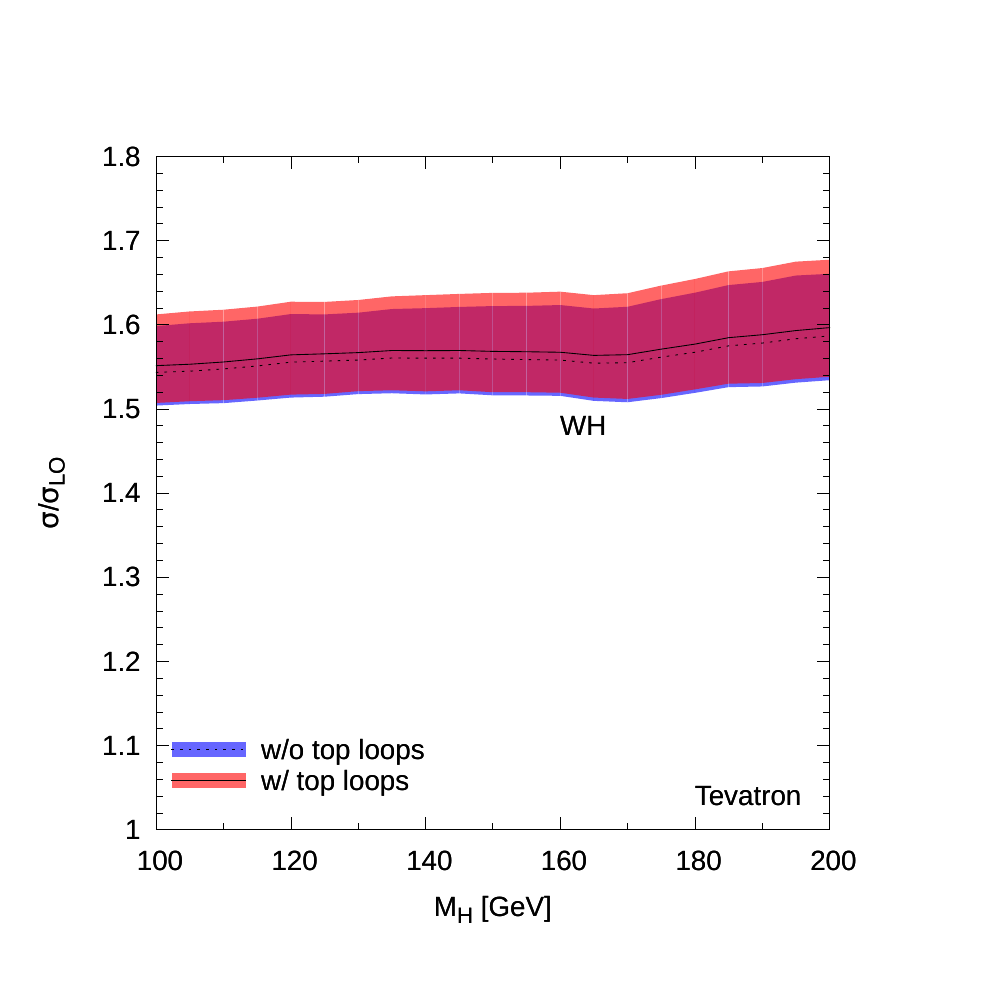} &
      \includegraphics[width=.45\textwidth]{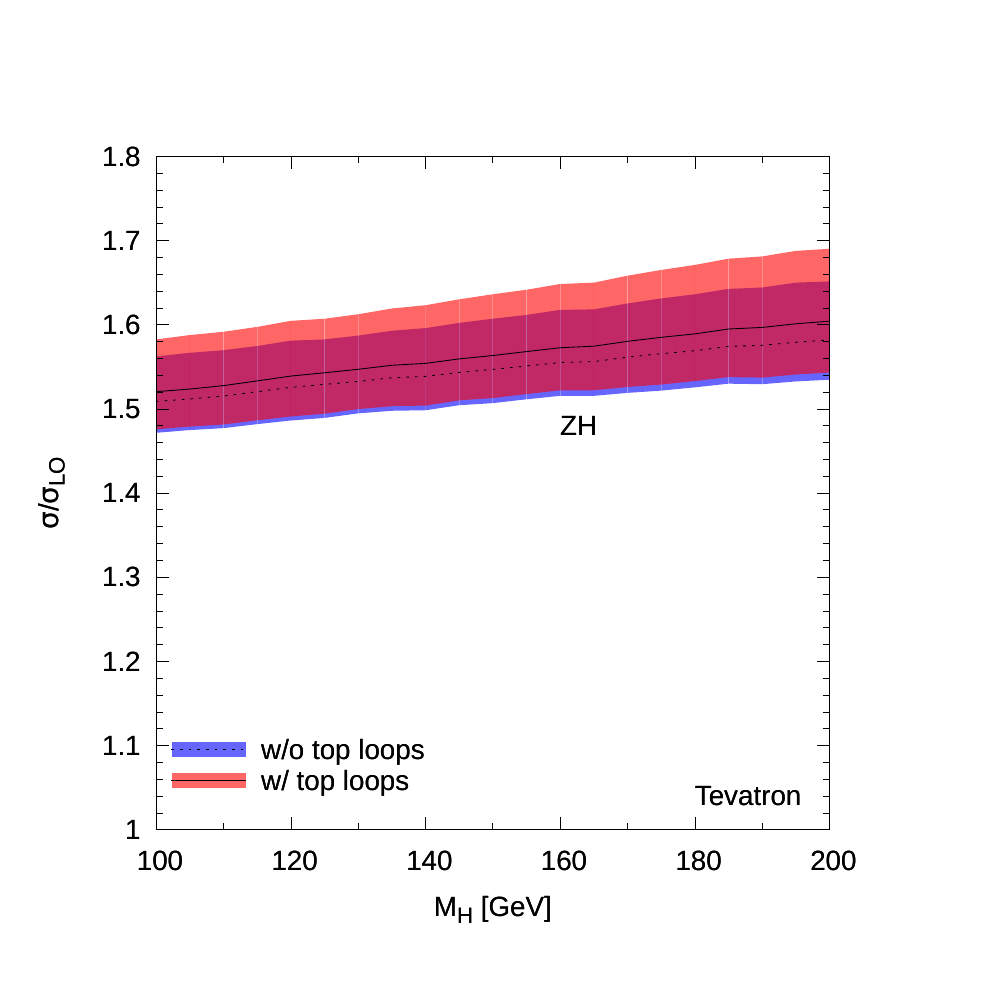}
      \\      (a) & (b)
    \end{tabular}
    \parbox{.9\textwidth}{
      \caption[]{\label{fig::kfactevatron}\sloppy K-factor at the Tevatron
        for (a) $\hw{}$ and (b) $\hz{}$ production, including \nnlo{}
        \qcd{} and electro-weak corrections, with and without the newly
        evaluated top-quark induced terms.  }}
  \end{center}
\end{figure}

Finally, Figs.\,\ref{fig::kfaclhc} and \ref{fig::kfactevatron} show the
K-factors resulting from our compilation, including \nnlo{} \qcd{},
electro-weak effects, and the newly evaluated top-mediated terms of this
paper. In the \lhc{} case, we also show the previous numbers from
Ref.~\cite{Dittmaier:2011ti}, and for the Tevatron we provide the
analogous numbers with the same input data.

These plots clearly show that for the \lhc{}, the top-mediated terms
affect the $\hw{}$ production cross section at the order of the
estimated perturbative uncertainties. In fact, note that for
$\sqrt{s}=14$\,TeV and above around $\mhiggs=160$\,GeV, the uncertainty
bands of the predictions with and without the top-induced terms hardly
overlap. Since the uncertainty bands of the $\hz{}$ channel are
significantly larger than for $\hw{}$, mostly due to the $gg$-induced
terms of \fig{fig::logg}\,(c), the relative importance of the
top-mediated terms on $\hz{}$ is much smaller than for $\hw{}$.

At the Tevatron, the K-factor is typically larger than at the \lhc{},
and the top-mediated terms have a rather small impact. Also, the
difference between $\hw{}$ and $\hz{}$ in the K-factor is much less
pronounced than at the \lhc{}, the reason being again the $gg$-induced
terms which have much smaller impact at the Tevatron.


%
%


%
%




\section{Conclusions and Outlook}

The effects of an as of yet neglected contribution to the
\higgsstrahlung{} process at hadron colliders have been studied. The
relevant Feynman diagrams contain closed top-quark loops and are of
order $g^3\lambda_t\alpha_s^2$. The one-loop real-emission contributions
were evaluated exactly, while the two-loop virtual terms were evaluated
in the heavy-top limit. It was argued that this provides a reliable
approximation to the exact result. The numerical impact of the newly
evaluated terms is within the current estimate for the theoretical
uncertainty, but may well become important once the uncertainty induced
by \pdf{}s and the strong coupling reduces.

We plan to include the newly evaluated terms in the publicly available
program {\tt vh@nnlo} in order to provide a tool for the evaluation of
the the full $\order{\alpha_s^2}$ prediction for $\hz{}$ and $\hw{}$
production.

Finally, let us remark that similar contributions exist also for the
weak-boson fusion
process~\cite{Rainwater:1997dg,Han:1992hr,Harlander:2008xn,Bolzoni:2010xr}.
Parts of them were evaluated in Ref.~\cite{Bolzoni:2011cu}.


\paragraph{Acknowledgments.}

We would like to thank Stefan Dittmaier and Abdelhak Djouadi for
discussions, and Hendrik Mantler for helpful input. RVH and MW thank the
{\abbrev CERN PH-TH} group for hospitality.  This work was supported by
{\it Deutsche Forschungsgemeinschaft}, contract HA~2990/5-1, the Helmholtz
Alliance {\it Physics at the Terascale}, and the Initial Training
Network {\it LHCPhenoNet}.




\def\app#1#2#3{{\it Act.~Phys.~Pol.~}\jref{\bf B #1}{#2}{#3}}
\def\apa#1#2#3{{\it Act.~Phys.~Austr.~}\jref{\bf#1}{#2}{#3}}
\def\annphys#1#2#3{{\it Ann.~Phys.~}\jref{\bf #1}{#2}{#3}}
\def\cmp#1#2#3{{\it Comm.~Math.~Phys.~}\jref{\bf #1}{#2}{#3}}
\def\cpc#1#2#3{{\it Comp.~Phys.~Commun.~}\jref{\bf #1}{#2}{#3}}
\def\epjc#1#2#3{{\it Eur.\ Phys.\ J.\ }\jref{\bf C #1}{#2}{#3}}
\def\fortp#1#2#3{{\it Fortschr.~Phys.~}\jref{\bf#1}{#2}{#3}}
\def\ijmpc#1#2#3{{\it Int.~J.~Mod.~Phys.~}\jref{\bf C #1}{#2}{#3}}
\def\ijmpa#1#2#3{{\it Int.~J.~Mod.~Phys.~}\jref{\bf A #1}{#2}{#3}}
\def\jcp#1#2#3{{\it J.~Comp.~Phys.~}\jref{\bf #1}{#2}{#3}}
\def\jetp#1#2#3{{\it JETP~Lett.~}\jref{\bf #1}{#2}{#3}}
\def\jphysg#1#2#3{{\small\it J.~Phys.~G~}\jref{\bf #1}{#2}{#3}}
\def\jhep#1#2#3{{\small\it JHEP~}\jref{\bf #1}{#2}{#3}}
\def\mpl#1#2#3{{\it Mod.~Phys.~Lett.~}\jref{\bf A #1}{#2}{#3}}
\def\nima#1#2#3{{\it Nucl.~Inst.~Meth.~}\jref{\bf A #1}{#2}{#3}}
\def\npb#1#2#3{{\it Nucl.~Phys.~}\jref{\bf B #1}{#2}{#3}}
\def\nca#1#2#3{{\it Nuovo~Cim.~}\jref{\bf #1A}{#2}{#3}}
\def\plb#1#2#3{{\it Phys.~Lett.~}\jref{\bf B #1}{#2}{#3}}
\def\prc#1#2#3{{\it Phys.~Reports }\jref{\bf #1}{#2}{#3}}
\def\prd#1#2#3{{\it Phys.~Rev.~}\jref{\bf D #1}{#2}{#3}}
\def\pR#1#2#3{{\it Phys.~Rev.~}\jref{\bf #1}{#2}{#3}}
\def\prl#1#2#3{{\it Phys.~Rev.~Lett.~}\jref{\bf #1}{#2}{#3}}
\def\pr#1#2#3{{\it Phys.~Reports }\jref{\bf #1}{#2}{#3}}
\def\ptp#1#2#3{{\it Prog.~Theor.~Phys.~}\jref{\bf #1}{#2}{#3}}
\def\ppnp#1#2#3{{\it Prog.~Part.~Nucl.~Phys.~}\jref{\bf #1}{#2}{#3}}
\def\rmp#1#2#3{{\it Rev.~Mod.~Phys.~}\jref{\bf #1}{#2}{#3}}
\def\sovnp#1#2#3{{\it Sov.~J.~Nucl.~Phys.~}\jref{\bf #1}{#2}{#3}}
\def\sovus#1#2#3{{\it Sov.~Phys.~Usp.~}\jref{\bf #1}{#2}{#3}}
\def\tmf#1#2#3{{\it Teor.~Mat.~Fiz.~}\jref{\bf #1}{#2}{#3}}
\def\tmp#1#2#3{{\it Theor.~Math.~Phys.~}\jref{\bf #1}{#2}{#3}}
\def\yadfiz#1#2#3{{\it Yad.~Fiz.~}\jref{\bf #1}{#2}{#3}}
\def\zpc#1#2#3{{\it Z.~Phys.~}\jref{\bf C #1}{#2}{#3}}
\def\ibid#1#2#3{{ibid.~}\jref{\bf #1}{#2}{#3}}
\def\otherjournal#1#2#3#4{{\it #1}\jref{\bf #2}{#3}{#4}}

\newcommand{\jref}[3]{{\bf #1} (#2) #3}
\newcommand{\hepph}[1]{{\tt hep-ph/#1}}
\newcommand{\hepth}[1]{{\tt hep-th/#1}}
\newcommand{\mathph}[1]{{\tt math-ph/#1}}
\newcommand{\arxiv}[2]{{\tt arXiv:#1}}
\newcommand{\bibentry}[4]{#1, {\it #2}, #3\ifthenelse{\equal{#4}{}}{}{, }#4.}


\end{document}

%% file: inputs/wzh7nnlo-fb.tex
\begin{tabular}{|c||c|c|c||c|c|c|}
\hline
$\mhiggs$ & $\sigma(\hw)$ & Pert & \pdf+$\alpha_s$ & $\sigma(\hz)$ & Pert & \pdf+$\alpha_s$ \\\footnotesize[GeV] & \footnotesize[fb] & \footnotesize[\%] &\footnotesize[\%] & \footnotesize[fb] & \footnotesize[\%] &\footnotesize[\%] \\\hline &&&&&&\\[-1em]
$ 100 $ & $ 1197 $ & $^{ + 1.0 }_{ -0.8 } $ & 
$ \pm 3.4 $
 & $    636.8 $ & $^{ + 1.5 }_{ -1.5 } $ &
$ \pm 3.4 $ \\[.2em]

$ 105 $ & $ 1027 $ & $^{ + 0.8 }_{ -1.1 } $ & 
$ \pm 3.5 $
 & $    549.8 $ & $^{ + 1.7 }_{ -1.9 } $ &
$ \pm 3.7 $ \\[.2em]

$ 110 $ & $    883.7 $ & $^{ + 0.7 }_{ -1.0 } $ & 
$ \pm 3.8 $
 & $    476.5 $ & $^{ + 1.6 }_{ -1.6 } $ &
$ \pm 4.1 $ \\[.2em]

$ 115 $ & $    762.0 $ & $^{ + 0.8 }_{ -1.1 } $ & 
$ \pm 3.9 $
 & $    414.6 $ & $^{ + 1.7 }_{ -1.5 } $ &
$ \pm 4.2 $ \\[.2em]

$ 120 $ & $    662.7 $ & $^{ + 0.8 }_{ -1.1 } $ & 
$ \pm 3.4 $
 & $    363.3 $ & $^{ + 2.0 }_{ -1.6 } $ &
$ \pm 3.5 $ \\[.2em]

$ 125 $ & $    578.8 $ & $^{ + 0.6 }_{ -1.2 } $ & 
$ \pm 3.5 $
 & $    319.0 $ & $^{ + 1.9 }_{ -2.0 } $ &
$ \pm 3.5 $ \\[.2em]

$ 130 $ & $    506.1 $ & $^{ + 0.8 }_{ -1.1 } $ & 
$ \pm 3.5 $
 & $    280.7 $ & $^{ + 2.0 }_{ -1.8 } $ &
$ \pm 3.7 $ \\[.2em]

$ 135 $ & $    443.7 $ & $^{ + 1.1 }_{ -0.8 } $ & 
$ \pm 3.4 $
 & $    247.9 $ & $^{ + 2.2 }_{ -1.8 } $ &
$ \pm 3.6 $ \\[.2em]

$ 140 $ & $    389.9 $ & $^{ + 1.0 }_{ -0.9 } $ & 
$ \pm 3.5 $
 & $    219.5 $ & $^{ + 2.0 }_{ -2.0 } $ &
$ \pm 3.7 $ \\[.2em]

$ 145 $ & $    344.4 $ & $^{ + 0.7 }_{ -1.1 } $ & 
$ \pm 3.8 $
 & $    195.2 $ & $^{ + 2.3 }_{ -2.2 } $ &
$ \pm 4.0 $ \\[.2em]

$ 150 $ & $    303.5 $ & $^{ + 0.9 }_{ -1.1 } $ & 
$ \pm 3.3 $
 & $    173.2 $ & $^{ + 2.3 }_{ -2.0 } $ &
$ \pm 3.6 $ \\[.2em]

$ 155 $ & $    267.7 $ & $^{ + 1.0 }_{ -1.1 } $ & 
$ \pm 3.5 $
 & $    154.3 $ & $^{ + 2.6 }_{ -2.0 } $ &
$ \pm 3.6 $ \\[.2em]

$ 160 $ & $    231.9 $ & $^{ + 1.0 }_{ -1.1 } $ & 
$ \pm 3.8 $
 & $    135.0 $ & $^{ + 2.5 }_{ -2.1 } $ &
$ \pm 4.0 $ \\[.2em]

$ 165 $ & $    213.2 $ & $^{ + 1.0 }_{ -1.1 } $ & 
$ \pm 3.6 $
 & $    124.8 $ & $^{ + 2.7 }_{ -2.1 } $ &
$ \pm 4.1 $ \\[.2em]

$ 170 $ & $    190.7 $ & $^{ + 1.0 }_{ -1.1 } $ & 
$ \pm 3.8 $
 & $    112.0 $ & $^{ + 2.7 }_{ -2.3 } $ &
$ \pm 4.2 $ \\[.2em]

$ 175 $ & $    171.0 $ & $^{ + 0.8 }_{ -1.5 } $ & 
$ \pm 3.8 $
 & $    100.8 $ & $^{ + 2.7 }_{ -2.3 } $ &
$ \pm 4.1 $ \\[.2em]

$ 180 $ & $    154.0 $ & $^{ + 1.1 }_{ -1.0 } $ & 
$ \pm 3.5 $
 & $    90.34 $ & $^{ + 2.8 }_{ -2.3 } $ &
$ \pm 3.8 $ \\[.2em]

$ 185 $ & $    140.5 $ & $^{ + 0.9 }_{ -1.2 } $ & 
$ \pm 3.5 $
 & $    82.46 $ & $^{ + 2.9 }_{ -2.4 } $ &
$ \pm 3.8 $ \\[.2em]

$ 190 $ & $    126.9 $ & $^{ + 1.1 }_{ -1.1 } $ & 
$ \pm 3.7 $
 & $    74.65 $ & $^{ + 2.8 }_{ -2.6 } $ &
$ \pm 3.9 $ \\[.2em]

$ 195 $ & $    115.2 $ & $^{ + 1.2 }_{ -1.0 } $ & 
$ \pm 3.7 $
 & $    67.91 $ & $^{ + 2.9 }_{ -2.4 } $ &
$ \pm 4.0 $ \\[.2em]

$ 200 $ & $    104.6 $ & $^{ + 0.9 }_{ -1.3 } $ & 
$ \pm 3.8 $
 & $    61.81 $ & $^{ + 2.9 }_{ -2.4 } $ &
$ \pm 4.1 $ \\[.2em]

$ 210 $ & $    86.70 $ & $^{ + 1.0 }_{ -1.1 } $ & 
$ \pm 3.7 $
 & $    51.41 $ & $^{ + 2.8 }_{ -2.5 } $ &
$ \pm 4.2 $ \\[.2em]

$ 220 $ & $    72.38 $ & $^{ + 0.9 }_{ -1.3 } $ & 
$ \pm 3.7 $
 & $    42.97 $ & $^{ + 2.9 }_{ -2.4 } $ &
$ \pm 4.2 $ \\[.2em]

$ 230 $ & $    60.87 $ & $^{ + 1.3 }_{ -1.1 } $ & 
$ \pm 4.5 $
 & $    36.15 $ & $^{ + 2.8 }_{ -2.4 } $ &
$ \pm 4.8 $ \\[.2em]

$ 240 $ & $    51.45 $ & $^{ + 1.1 }_{ -1.2 } $ & 
$ \pm 4.0 $
 & $    30.46 $ & $^{ + 2.6 }_{ -2.4 } $ &
$ \pm 4.4 $ \\[.2em]

$ 250 $ & $    43.68 $ & $^{ + 1.1 }_{ -1.1 } $ & 
$ \pm 4.0 $
 & $    25.81 $ & $^{ + 2.7 }_{ -2.2 } $ &
$ \pm 4.2 $ \\[.2em]

$ 260 $ & $    37.25 $ & $^{ + 1.3 }_{ -1.1 } $ & 
$ \pm 4.0 $
 & $    21.94 $ & $^{ + 2.6 }_{ -2.3 } $ &
$ \pm 4.5 $ \\[.2em]

$ 270 $ & $    31.91 $ & $^{ + 1.2 }_{ -1.3 } $ & 
$ \pm 3.8 $
 & $    18.70 $ & $^{ + 2.4 }_{ -2.2 } $ &
$ \pm 4.3 $ \\[.2em]

$ 280 $ & $    27.39 $ & $^{ + 0.9 }_{ -1.4 } $ & 
$ \pm 4.4 $
 & $    16.02 $ & $^{ + 2.4 }_{ -1.9 } $ &
$ \pm 4.9 $ \\[.2em]

$ 290 $ & $    23.65 $ & $^{ + 1.2 }_{ -1.2 } $ & 
$ \pm 4.2 $
 & $    13.79 $ & $^{ + 2.3 }_{ -2.0 } $ &
$ \pm 4.5 $ \\[.2em]

$ 300 $ & $    20.47 $ & $^{ + 1.1 }_{ -1.3 } $ & 
$ \pm 4.5 $
 & $    11.90 $ & $^{ + 2.2 }_{ -1.8 } $ &
$ \pm 5.0 $ \\[.2em]

\hline
\end{tabular}

%% file: inputs/wzh14nnlo-fb.tex
\begin{tabular}{|c||c|c|c||c|c|c|}
\hline
$\mhiggs$ & $\sigma(\hw)$ & Pert & \pdf+$\alpha_s$ & $\sigma(\hz)$ & Pert & \pdf+$\alpha_s$ \\\footnotesize[GeV] & \footnotesize[fb] & \footnotesize[\%] &\footnotesize[\%] & \footnotesize[fb] & \footnotesize[\%] &\footnotesize[\%] \\\hline &&&&&&\\[-1em]
$ 100 $ & $ 3035 $ & $^{ + 1.3 }_{ -0.9 } $ & 
$ \pm 3.7 $
 & $ 1700 $ & $^{ + 2.3 }_{ -1.9 } $ &
$ \pm 3.8 $ \\[.2em]

$ 105 $ & $ 2625 $ & $^{ + 1.1 }_{ -0.8 } $ & 
$ \pm 3.5 $
 & $ 1483 $ & $^{ + 2.2 }_{ -2.1 } $ &
$ \pm 3.7 $ \\[.2em]

$ 110 $ & $ 2273 $ & $^{ + 0.8 }_{ -1.1 } $ & 
$ \pm 3.8 $
 & $ 1297 $ & $^{ + 2.6 }_{ -2.0 } $ &
$ \pm 4.0 $ \\[.2em]

$ 115 $ & $ 1976 $ & $^{ + 1.2 }_{ -0.6 } $ & 
$ \pm 3.8 $
 & $ 1142 $ & $^{ + 2.9 }_{ -1.9 } $ &
$ \pm 3.7 $ \\[.2em]

$ 120 $ & $ 1731 $ & $^{ + 1.1 }_{ -0.7 } $ & 
$ \pm 3.8 $
 & $ 1008 $ & $^{ + 2.9 }_{ -2.2 } $ &
$ \pm 3.6 $ \\[.2em]

$ 125 $ & $ 1523 $ & $^{ + 0.9 }_{ -1.0 } $ & 
$ \pm 3.8 $
 & $    893.2 $ & $^{ + 3.2 }_{ -2.2 } $ &
$ \pm 3.7 $ \\[.2em]

$ 130 $ & $ 1342 $ & $^{ + 1.0 }_{ -0.8 } $ & 
$ \pm 3.3 $
 & $    793.9 $ & $^{ + 3.4 }_{ -2.2 } $ &
$ \pm 3.4 $ \\[.2em]

$ 135 $ & $ 1183 $ & $^{ + 1.2 }_{ -0.9 } $ & 
$ \pm 2.9 $
 & $    706.6 $ & $^{ + 3.4 }_{ -2.6 } $ &
$ \pm 3.0 $ \\[.2em]

$ 140 $ & $ 1048 $ & $^{ + 0.8 }_{ -1.1 } $ & 
$ \pm 3.1 $
 & $    633.4 $ & $^{ + 3.4 }_{ -2.6 } $ &
$ \pm 3.0 $ \\[.2em]

$ 145 $ & $    933.1 $ & $^{ + 1.1 }_{ -0.8 } $ & 
$ \pm 3.3 $
 & $    567.2 $ & $^{ + 3.8 }_{ -2.5 } $ &
$ \pm 3.4 $ \\[.2em]

$ 150 $ & $    827.5 $ & $^{ + 0.9 }_{ -1.1 } $ & 
$ \pm 2.7 $
 & $    508.2 $ & $^{ + 3.8 }_{ -2.4 } $ &
$ \pm 2.7 $ \\[.2em]

$ 155 $ & $    736.3 $ & $^{ + 0.9 }_{ -1.0 } $ & 
$ \pm 3.1 $
 & $    457.3 $ & $^{ + 3.8 }_{ -2.8 } $ &
$ \pm 3.2 $ \\[.2em]

$ 160 $ & $    644.0 $ & $^{ + 0.8 }_{ -0.9 } $ & 
$ \pm 3.1 $
 & $    404.1 $ & $^{ + 4.1 }_{ -2.8 } $ &
$ \pm 3.1 $ \\[.2em]

$ 165 $ & $    594.0 $ & $^{ + 0.8 }_{ -1.0 } $ & 
$ \pm 2.4 $
 & $    375.6 $ & $^{ + 4.3 }_{ -2.7 } $ &
$ \pm 2.6 $ \\[.2em]

$ 170 $ & $    534.3 $ & $^{ + 0.9 }_{ -1.1 } $ & 
$ \pm 2.8 $
 & $    340.2 $ & $^{ + 4.1 }_{ -2.9 } $ &
$ \pm 3.0 $ \\[.2em]

$ 175 $ & $    483.9 $ & $^{ + 1.1 }_{ -0.8 } $ & 
$ \pm 2.9 $
 & $    308.8 $ & $^{ + 4.0 }_{ -3.1 } $ &
$ \pm 3.1 $ \\[.2em]

$ 180 $ & $    434.5 $ & $^{ + 1.1 }_{ -1.0 } $ & 
$ \pm 2.8 $
 & $    278.5 $ & $^{ + 4.3 }_{ -3.3 } $ &
$ \pm 3.0 $ \\[.2em]

$ 185 $ & $    402.8 $ & $^{ + 1.0 }_{ -1.1 } $ & 
$ \pm 2.5 $
 & $    256.2 $ & $^{ + 4.1 }_{ -3.3 } $ &
$ \pm 2.6 $ \\[.2em]

$ 190 $ & $    366.0 $ & $^{ + 0.9 }_{ -1.1 } $ & 
$ \pm 2.8 $
 & $    233.6 $ & $^{ + 4.1 }_{ -3.3 } $ &
$ \pm 3.0 $ \\[.2em]

$ 195 $ & $    334.6 $ & $^{ + 0.9 }_{ -1.2 } $ & 
$ \pm 2.7 $
 & $    214.4 $ & $^{ + 4.1 }_{ -3.4 } $ &
$ \pm 2.9 $ \\[.2em]

$ 200 $ & $    305.5 $ & $^{ + 1.0 }_{ -1.0 } $ & 
$ \pm 3.0 $
 & $    196.6 $ & $^{ + 4.2 }_{ -3.5 } $ &
$ \pm 3.1 $ \\[.2em]

$ 210 $ & $    257.0 $ & $^{ + 0.9 }_{ -1.1 } $ & 
$ \pm 2.6 $
 & $    165.4 $ & $^{ + 4.5 }_{ -2.9 } $ &
$ \pm 2.6 $ \\[.2em]

$ 220 $ & $    217.5 $ & $^{ + 1.3 }_{ -1.0 } $ & 
$ \pm 2.8 $
 & $    140.3 $ & $^{ + 4.1 }_{ -3.2 } $ &
$ \pm 2.9 $ \\[.2em]

$ 230 $ & $    185.9 $ & $^{ + 1.1 }_{ -0.9 } $ & 
$ \pm 3.5 $
 & $    119.3 $ & $^{ + 4.1 }_{ -3.1 } $ &
$ \pm 3.6 $ \\[.2em]

$ 240 $ & $    158.9 $ & $^{ + 1.1 }_{ -1.0 } $ & 
$ \pm 3.3 $
 & $    101.7 $ & $^{ + 3.9 }_{ -3.0 } $ &
$ \pm 3.4 $ \\[.2em]

$ 250 $ & $    136.8 $ & $^{ + 0.8 }_{ -1.2 } $ & 
$ \pm 3.0 $
 & $    86.96 $ & $^{ + 3.7 }_{ -2.9 } $ &
$ \pm 3.2 $ \\[.2em]

$ 260 $ & $    118.3 $ & $^{ + 0.9 }_{ -1.2 } $ & 
$ \pm 2.8 $
 & $    74.80 $ & $^{ + 3.8 }_{ -2.6 } $ &
$ \pm 3.1 $ \\[.2em]

$ 270 $ & $    102.8 $ & $^{ + 1.2 }_{ -1.1 } $ & 
$ \pm 2.6 $
 & $    64.48 $ & $^{ + 3.3 }_{ -2.5 } $ &
$ \pm 2.8 $ \\[.2em]

$ 280 $ & $    89.49 $ & $^{ + 1.1 }_{ -1.1 } $ & 
$ \pm 3.0 $
 & $    55.83 $ & $^{ + 3.3 }_{ -2.4 } $ &
$ \pm 3.2 $ \\[.2em]

$ 290 $ & $    78.61 $ & $^{ + 1.0 }_{ -1.1 } $ & 
$ \pm 3.2 $
 & $    48.67 $ & $^{ + 3.0 }_{ -2.2 } $ &
$ \pm 3.2 $ \\[.2em]

$ 300 $ & $    68.86 $ & $^{ + 1.3 }_{ -1.1 } $ & 
$ \pm 3.3 $
 & $    42.43 $ & $^{ + 2.9 }_{ -2.2 } $ &
$ \pm 3.6 $ \\[.2em]

\hline\end{tabular}

%% file: inputs/wzhtevatronnnlo-fb.tex
\begin{tabular}{|c||c|c|c||c|c|c|}
\hline
$\mhiggs$ & $\sigma(\hw)$ & Pert & \pdf+$\alpha_s$ & $\sigma(\hz)$ & Pert & \pdf+$\alpha_s$ \\\footnotesize[GeV] & \footnotesize[fb] & \footnotesize[\%] &\footnotesize[\%] & \footnotesize[fb] & \footnotesize[\%] &\footnotesize[\%] \\\hline &&&&&&\\[-1em]
$ 100 $ & $    278.0 $ & $^{ + 4.0 }_{ -2.9 } $ & 
$ \pm 5.1 $
 & $    161.2 $ & $^{ + 4.1 }_{ -3.0 } $ &
$ \pm 5.5 $ \\[.2em]

$ 105 $ & $    235.7 $ & $^{ + 4.1 }_{ -2.8 } $ & 
$ \pm 5.3 $
 & $    138.2 $ & $^{ + 4.3 }_{ -3.0 } $ &
$ \pm 5.6 $ \\[.2em]

$ 110 $ & $    200.9 $ & $^{ + 4.0 }_{ -2.9 } $ & 
$ \pm 5.5 $
 & $    119.0 $ & $^{ + 4.2 }_{ -3.1 } $ &
$ \pm 5.7 $ \\[.2em]

$ 115 $ & $    172.1 $ & $^{ + 4.0 }_{ -3.0 } $ & 
$ \pm 5.5 $
 & $    103.0 $ & $^{ + 4.2 }_{ -3.1 } $ &
$ \pm 5.8 $ \\[.2em]

$ 120 $ & $    148.0 $ & $^{ + 4.1 }_{ -3.0 } $ & 
$ \pm 5.8 $
 & $    89.46 $ & $^{ + 4.3 }_{ -3.2 } $ &
$ \pm 5.9 $ \\[.2em]

$ 125 $ & $    127.5 $ & $^{ + 4.0 }_{ -3.1 } $ & 
$ \pm 6.0 $
 & $    77.86 $ & $^{ + 4.2 }_{ -3.2 } $ &
$ \pm 6.1 $ \\[.2em]

$ 130 $ & $    110.2 $ & $^{ + 4.0 }_{ -2.9 } $ & 
$ \pm 6.2 $
 & $    67.98 $ & $^{ + 4.3 }_{ -3.1 } $ &
$ \pm 6.0 $ \\[.2em]

$ 135 $ & $    95.61 $ & $^{ + 4.1 }_{ -3.0 } $ & 
$ \pm 6.1 $
 & $    59.55 $ & $^{ + 4.4 }_{ -3.2 } $ &
$ \pm 6.0 $ \\[.2em]

$ 140 $ & $    83.06 $ & $^{ + 4.2 }_{ -3.1 } $ & 
$ \pm 6.4 $
 & $    52.23 $ & $^{ + 4.5 }_{ -3.3 } $ &
$ \pm 6.2 $ \\[.2em]

$ 145 $ & $    72.36 $ & $^{ + 4.3 }_{ -3.0 } $ & 
$ \pm 6.5 $
 & $    46.01 $ & $^{ + 4.6 }_{ -3.2 } $ &
$ \pm 6.3 $ \\[.2em]

$ 150 $ & $    63.16 $ & $^{ + 4.4 }_{ -3.1 } $ & 
$ \pm 6.3 $
 & $    40.59 $ & $^{ + 4.7 }_{ -3.3 } $ &
$ \pm 6.0 $ \\[.2em]

$ 155 $ & $    55.28 $ & $^{ + 4.4 }_{ -3.1 } $ & 
$ \pm 6.9 $
 & $    35.91 $ & $^{ + 4.8 }_{ -3.3 } $ &
$ \pm 6.6 $ \\[.2em]

$ 160 $ & $    48.49 $ & $^{ + 4.6 }_{ -3.0 } $ & 
$ \pm 6.9 $
 & $    31.84 $ & $^{ + 4.9 }_{ -3.3 } $ &
$ \pm 6.5 $ \\[.2em]

$ 165 $ & $    42.54 $ & $^{ + 4.5 }_{ -3.2 } $ & 
$ \pm 7.8 $
 & $    28.23 $ & $^{ + 4.9 }_{ -3.4 } $ &
$ \pm 7.3 $ \\[.2em]

$ 170 $ & $    37.51 $ & $^{ + 4.6 }_{ -3.3 } $ & 
$ \pm 7.1 $
 & $    25.15 $ & $^{ + 5.0 }_{ -3.5 } $ &
$ \pm 6.6 $ \\[.2em]

$ 175 $ & $    33.26 $ & $^{ + 4.7 }_{ -3.4 } $ & 
$ \pm 6.9 $
 & $    22.42 $ & $^{ + 5.2 }_{ -3.6 } $ &
$ \pm 6.4 $ \\[.2em]

$ 180 $ & $    29.53 $ & $^{ + 4.9 }_{ -3.4 } $ & 
$ \pm 7.3 $
 & $    20.02 $ & $^{ + 5.3 }_{ -3.6 } $ &
$ \pm 6.6 $ \\[.2em]

$ 185 $ & $    26.29 $ & $^{ + 4.9 }_{ -3.4 } $ & 
$ \pm 7.7 $
 & $    17.92 $ & $^{ + 5.4 }_{ -3.7 } $ &
$ \pm 6.9 $ \\[.2em]

$ 190 $ & $    23.39 $ & $^{ + 5.0 }_{ -3.6 } $ & 
$ \pm 7.9 $
 & $    16.03 $ & $^{ + 5.4 }_{ -3.8 } $ &
$ \pm 7.0 $ \\[.2em]

$ 195 $ & $    20.86 $ & $^{ + 5.1 }_{ -3.6 } $ & 
$ \pm 7.7 $
 & $    14.38 $ & $^{ + 5.6 }_{ -3.9 } $ &
$ \pm 6.8 $ \\[.2em]

$ 200 $ & $    18.61 $ & $^{ + 5.0 }_{ -3.6 } $ & 
$ \pm 7.6 $
 & $    12.91 $ & $^{ + 5.5 }_{ -3.9 } $ &
$ \pm 6.5 $ \\[.2em]

\hline
\end{tabular}